\documentclass{article}

\usepackage{amsmath, amsfonts, amssymb}
\usepackage{booktabs}
\usepackage{enumerate}
\usepackage{graphicx}
\usepackage{multirow}
\usepackage{bm}
\usepackage{amsthm}
\usepackage{listings}
\usepackage{bbm}
\usepackage[square, numbers, sort&compress]{natbib}
\usepackage{hyperref}

\newcommand*{\E}{\mathbbm{E}}
\newcommand*{\Var}{\mathrm{Var}}
\newcommand*{\Cov}{\mathrm{Cov}}

\renewcommand*{\vec}[1]{\boldsymbol{#1}}
\newcommand*{\mat}[1]{\mathrm{#1}}

\theoremstyle{plain}
\newtheorem{lem}{Lemma}

\theoremstyle{definition}

\begin{document}

\title{Kalman Filtering and Smoothing for Improving Precision in Horvitz--Thompson Estimation of Infectious Disease Prevalence}

\author{Jeongjin Lee$^{1}$, Grzegorz A.\ Rempala$^{1}$, Patrick M.\ Schnell$^{1, 2}$}

\date{\small
$^{1}$ Division of Biostatistics, College of Public Health, The Ohio State University \\
$^{2}$ Department of Immunology, Genetics and Pathology, Uppsala University
}

\maketitle
\sloppy

\begin{abstract}
Horvitz--Thompson (HT) estimators can provide unbiased daily estimates of infectious disease prevalence under repeated surveillance by correcting for nonrandom testing induced by scheduled, symptom-based, and contact-tracing components. 
However, because each HT estimate is based on the testing data available for that day and may involve highly variable inverse probability weights, it can be noisy, have precision that varies over time, and become unavailable during temporary interruptions in testing.
The daily HT estimator is modeled as a noisy observation of an underlying prevalence process, with day-specific observation variances estimated using a delete-a-group jackknife.
Our primary specification is a joint local linear trend state-space model that extends the standard level-only random walk by adding a latent slope. 
The Kalman filter improves precision by borrowing information from past estimates.
At each time $t$, the process variances are estimated or carried forward using only observations available through time $t$, so the resulting filtered estimate is available in real time.
We also describe the corresponding Kalman smoother as a retrospective extension based on the full observed series.
When daily HT estimates are missing, the Kalman filter proceeds through prediction-only updates, whereas the corresponding smoother retrospectively reconstructs those periods using later observations. 
In simulation studies based on repeated testing designs, the joint Kalman filter substantially improves precision relative to the raw daily HT estimator while preserving the main temporal pattern of the true prevalence trajectory, and the corresponding smoother provides a more stable retrospective summary. 
In an application to data from The Ohio State University’s fall 2020 SARS-CoV-2 surveillance program, the joint Kalman filter and smoother provide prevalence estimates even on weekends and other days with no testing, when the HT estimator provides neither point nor interval estimates. 
Their model-based confidence intervals are also narrower than the HT intervals on observed days.

\end{abstract}

\section{Introduction}

During the COVID-19 pandemic, many institutions implemented longitudinal testing and isolation programs to monitor infection and mitigate transmission. 
Such programs were used in settings including colleges and universities \citep{school21, college21, chang2021repeat}, workplaces \citep{work22}, and professional sports organizations \citep{mba21}. 
Beyond their operational role in identifying and isolating infectious individuals, the resulting testing data were also used to estimate infection prevalence in the monitored population, and accurate prevalence estimation became important for institutional risk assessment and public health decision-making \citep{baker2022successful}. 
A simple and widely used estimator of prevalence is the \textit{test-positive rate} (TPR), defined as the proportion of positive tests among all tests conducted on a given day. 
However, under repeated testing, the TPR is generally a biased estimator of true prevalence, as discussed in detail by \citet{schnell2024overcoming, lee2026counterfactual}.
The main reason is that tests are not administered as a simple random sample from the full population. 
Instead, testing is often carried out under nonuniform schedules, such as once-per-week surveillance, and is supplemented by symptom-based and contact-tracing components, so individuals at higher infection risk may be tested at different rates than others. 
Consequently, the set of individuals tested on a given day is not representative of the full at-risk population, and the daily TPR may systematically differ from the true prevalence \citep{schnell2024overcoming, lee2026counterfactual}.

To address this issue, unbiased prevalence estimators have been developed that explicitly account for the repeated testing process. 
In particular, \citet{schnell2024overcoming} adapted the Horvitz--Thompson (HT) estimator \citep{horvitz1952generalization} to prevalence estimation under repeated scheduled testing. 
Their approach uses inverse probability weighting based on each individual's probability of being tested, thereby correcting the selection induced by the testing process and targeting the true prevalence. 
Building on that work, \citet{lee2026counterfactual} extended the framework to repeated testing systems with scheduled, symptomatic, and contact-tracing components through a causal framework \citep{hernan2020causal}. 
This framework formalizes the assumptions underlying prevalence identification, clarifies how the testing process induces selection bias, and supports unbiased estimation under more complex testing designs. 
In practice, however, even these estimators may remain highly variable from day to day. 
Because both the amount and composition of testing vary over time, the precision of the daily estimates can also differ substantially across days, motivating the use of a method that uses information available up to the current day to produce a more stable prevalence trajectory.

We instead take the daily HT estimator \citep{schnell2024overcoming, lee2026counterfactual} as the starting point and study how to convert a noisy sequence of day-specific estimates into a stable and interpretable prevalence trajectory. 
Kalman filtering is well suited to this problem because it combines temporal structure with day-specific observation uncertainty and handles days on which the daily prevalence estimate is unavailable in a principled way \citep{Kalman1960, Harvey1989, DurbinKoopman2012}.
In our implementation, the process variances at time $t$ are selected using all observations available through time $t$, so the filtered prevalence estimate is a real-time estimate.
The same state-space formulation also yields a Kalman smoother, which revises the estimated prevalence at time $t$ using later observations and is therefore useful for retrospective summaries once the full series has been observed \citep{Harvey1989, DurbinKoopman2012}.
Our approach is based on a local linear trend state-space model with latent prevalence and latent slope components. 
This formulation extends the standard level-only random walk model by allowing the filter to carry forward information about whether prevalence is currently increasing or decreasing.
Across the simulation studies, both the joint filtered and joint smoothed estimates substantially improve precision relative to the daily HT estimates, yielding markedly lower variability while preserving the main temporal pattern of the underlying prevalence trajectory.
The default model-based confidence intervals exhibit undercoverage on some days, whereas confidence intervals centered on convex-combination estimators often achieve coverage closer to the nominal 95\% level, although the corresponding estimators yield smaller reductions in RMSE than the joint estimators.

The remainder of the paper is organized as follows. 
Section 2 introduces the joint state-space model, the Kalman filter, the corresponding smoother, and innovation-likelihood estimation. 
Sections 3 and 4 present simulation and real-data analyses centered on the joint model. 
Section 5 concludes with a discussion.

\section{Kalman Filtering and Smoothing of Prevalence}
Let $\hat p_t$ denote a daily prevalence estimator computed from the observed testing data at time $t$.
Examples include the Horvitz--Thompson (HT) estimator developed in \citet{schnell2024overcoming, lee2026counterfactual}.
We model $\hat p_t$ as a noisy observation of the first component of a latent bivariate state
\[
\alpha_t=  
\begin{pmatrix}
p_t\\
v_t
\end{pmatrix},
\]
where $p_t$ is the latent prevalence at time $t$ and $v_t$ is the latent slope, representing the expected change in prevalence from time $t$ to time $t+1$.
Compared with modeling $p_t$ alone, the joint formulation can carry forward recent upward or downward trends, which is especially useful when prevalence changes rapidly or testing data are temporarily unavailable.
We consider filtered prevalence estimates for prospective surveillance and smoothed prevalence estimates for retrospective summaries.

\subsection{State Space Model}
Our primary specification is the joint local linear trend state-space model \citep{Harvey1989, DurbinKoopman2012}, defined by
\begin{align}
\alpha_t
&= \mat{F}\alpha_{t-1} + \vec{w}_t,
&
\vec{w}_t
&\sim N(\vec{0},\mat{Q}),
\label{eq:kf_state}
\\
\hat p_t
&= \mat{H}\alpha_t + e_t,
&
e_t
&\sim N(0,R_t),
\label{eq:kf_meas}
\end{align}
with
\[
\mat{F}=
\begin{pmatrix}
1 & 1\\
0 & 1
\end{pmatrix},
\qquad
\mat{H}=
\begin{pmatrix}
1 & 0
\end{pmatrix},
\qquad
\mat{Q}=
\begin{pmatrix}
Q_{\text{level}} & 0\\
0 & Q_{\text{slope}}
\end{pmatrix}.
\]
Equivalently,
\begin{align}
p_t &= p_{t-1} + v_{t-1} + \eta_t,
\\
v_t &= v_{t-1} + \zeta_t,
\\
\hat p_t &= p_t + e_t,
\end{align}
where $\eta_t$ and $\zeta_t$ are mutually independent Gaussian errors with variances $Q_{\text{level}}$ and $Q_{\text{slope}}$, respectively.
Thus, the joint model extends the level-only random walk by adding a stochastic local slope and reduces to the level-only specification when $Q_{\text{slope}}=0$ and the initial slope is fixed at zero.
The observation variance $R_t$ represents the day-specific variability of the HT estimator around the latent prevalence $p_t$.
In applications, it is estimated using a delete-a-group jackknife \citep{kott2001delete} or another resampling method \citep{efron1987better} because dependence induced by complex repeated-testing designs complicates analytic variance estimation.

For time points $t$ and $s$ in the analysis period, define
\begin{align}
\tilde{\alpha}_{t\mid s}
&=
\E[\alpha_t\mid \hat p_1,\dots,\hat p_s],
\label{eq:kf_state_mean_def}
\\
\mat{P}_{t\mid s}
&=
\Var[\alpha_t\mid \hat p_1,\dots,\hat p_s].
\label{eq:kf_state_var_def}
\end{align}
When some daily estimates are missing, conditioning on $\hat p_1,\dots,\hat p_s$ refers only to the estimates observed through time $s$.
Thus, $\tilde{\alpha}_{t\mid t-1}$ is the one-step-ahead predicted state mean at time $t$, $\tilde{\alpha}_{t\mid t}$ is the filtered state mean after incorporating $\hat p_t$, and $\tilde{\alpha}_{t\mid T}$ is the fixed-interval smoothed state mean based on the full series through time $T$.
We write
\[
\tilde{\alpha}_{t\mid s}
=
\begin{pmatrix}
\tilde p_{t\mid s}\\
\tilde v_{t\mid s}
\end{pmatrix},
\]
where $\tilde p_{t\mid s}$ is the estimated prevalence level at time $t$ based on observations through time $s$, and $\tilde v_{t\mid s}$ is the estimated slope at time $t$ based on the same information.

\subsection{Initialization}
\label{subsec:initialization}
In practice, the first usable prevalence estimate need not occur on day 1. 
Some early days may be excluded because the estimator is undefined, because too few tests were observed, or because the corresponding variance estimate is unavailable. 
We therefore initialize the filter at the first time point
\begin{equation}
t_0=\min\left\{t: \hat p_t\ \text{and}\ R_t\ \text{are observed}\right\}.
\label{eq:kf_t0}
\end{equation}
At $t_0$, we initialize the filtered state directly as
\[
\tilde{\alpha}_{t_0\mid t_0}
=
\begin{pmatrix}
\hat p_{t_0}\\
0
\end{pmatrix},
\qquad
\mat{P}_{t_0\mid t_0}
=
\begin{pmatrix}
R_{t_0} & 0\\
0 & Q_{\text{slope}}
\end{pmatrix}.
\]
Thus, the initial prevalence level is set to the first usable daily estimate, its variance is set to the corresponding observation variance, and the initial slope is centered at zero with variance $Q_{\text{slope}}$.
The forward recursion then begins at $t=t_0+1$.

\subsection{Kalman Filter Recursion}
The Kalman filter proceeds recursively through prediction and update steps. 
For $t=t_0+1,\dots,T$, the prediction step is
\begin{align}
\tilde{\alpha}_{t\mid t-1}
&=
\mat{F}\tilde{\alpha}_{t-1\mid t-1},
\label{eq:kf_pred_mean}
\\
\mat{P}_{t\mid t-1}
&=
\mat{F}\mat{P}_{t-1\mid t-1}\mat{F}^{\top}+\mat{Q}.
\label{eq:kf_pred_var}
\end{align}
In particular, the one-step-ahead prediction for the prevalence level is
\[
\tilde p_{t\mid t-1}
=
\tilde p_{t-1\mid t-1}
+
\tilde v_{t-1\mid t-1},    
\]
so the slope term lets the filter continue an upward or downward movement instead of simply repeating the previous prevalence level.

The innovation, defined as the difference between the observed daily estimate and its one-step-ahead prediction, is given by
\begin{equation}
\nu_t = \hat p_t-\mat{H}\tilde{\alpha}_{t\mid t-1}
=
\hat p_t-\tilde p_{t\mid t-1},
\label{eq:kf_innovation}
\end{equation}
with corresponding variance
\begin{equation}
S_t = \mat{H}\mat{P}_{t\mid t-1}\mat{H}^{\top}+R_t.
\label{eq:kf_innovation_var}
\end{equation}
The Kalman gain,
\begin{equation}
\vec{K}_t
=
\mat{P}_{t\mid t-1}\mat{H}^{\top}S_t^{-1},
\label{eq:kf_gain}
\end{equation}
weights the innovation when updating the predicted state. 
The corresponding update equations are
\begin{align}
\tilde{\alpha}_{t\mid t}
&=
\tilde{\alpha}_{t\mid t-1}
+
\vec{K}_t\nu_t,
\label{eq:kf_update_mean}
\\
\mat{P}_{t\mid t}
&=
\left(\mat{I}-\vec{K}_t\mat{H}\right)\mat{P}_{t\mid t-1}.
\label{eq:kf_update_var}
\end{align}
The filtered prevalence estimate is the first component of $\tilde{\alpha}_{t\mid t}$, namely $\tilde p_{t\mid t}$. 
Because $\mat{H}= \begin{pmatrix} 1 & 0 \end{pmatrix}$, the prevalence update can be written in scalar form as
\begin{equation}
\tilde p_{t\mid t}
=
(1-k_t)\tilde p_{t\mid t-1}
+
k_t\hat p_t,
\qquad
k_t
=
\frac{[\mat{P}_{t\mid t-1}]_{11}}{[\mat{P}_{t\mid t-1}]_{11}+R_t}.
\label{eq:kf_update_prevalence_scalar}
\end{equation}
Thus, $\tilde p_{t\mid t}$ is a weighted average of the one-step-ahead predicted prevalence and the observed daily HT estimate, with more weight placed on $\hat p_t$ when $R_t$ is small relative to the prediction variance and more weight placed on the prediction when $R_t$ is large.
These recursions are stated for a given $\mat{Q}$.
The time-specific procedure used to select $\mat{Q}$ is described in Section~\ref{subsec:process-variance-selection}.

Under the state-space model, $[\mat{P}_{t\mid t}]_{11}$ is the conditional variance of the latent prevalence level given observations through time $t$ and can be compared with the day-specific observation variance of the raw HT estimator, $R_t$. 
This comparison quantifies the reduction in conditional uncertainty produced by the Kalman update.
\begin{lem}
\label{lem:ht_filter_variance}
For any given $\mat{Q}$ and any day $t$ for which $\hat p_t$ is observed, the conditional variance of the latent prevalence level given observations through time $t$ is no greater than the corresponding HT observation variance:
\[
[\mat{P}_{t\mid t}]_{11}\le R_t.
\]
\end{lem}
The proof is given in the Appendix.

\subsection{Kalman Smoother Recursion}
Once the full observed series through time $T$ is available, the same fitted state-space model yields a fixed-interval smoother \citep{Harvey1989, DurbinKoopman2012}. 
The backward recursion is initialized at the final time point $T$ using the filtered quantities $\tilde{\alpha}_{T\mid T}$ and $\mat{P}_{T\mid T}$.
For $t=T-1,\dots,t_0$, define the smoothing gain
\begin{equation}
\mat{J}_t
=
\mat{P}_{t\mid t}\mat{F}^{\top}\mat{P}_{t+1\mid t}^{-1}.
\label{eq:ks_gain}
\end{equation}
This expression assumes that $\mat{P}_{t+1\mid t}$ is nonsingular. At the exact boundary $Q_{\text{slope}}=0$ under the initialization in Section~\ref{subsec:initialization}, the degenerate joint model is handled using its equivalent level-only representation.
The smoothed state mean and covariance then satisfy the backward recursions
\begin{align}
\tilde{\alpha}_{t\mid T}
&=
\tilde{\alpha}_{t\mid t}
+
\mat{J}_t\left(\tilde{\alpha}_{t+1\mid T}-\tilde{\alpha}_{t+1\mid t}\right),
\label{eq:ks_mean}
\\
\mat{P}_{t\mid T}
&=
\mat{P}_{t\mid t}
+
\mat{J}_t\left(\mat{P}_{t+1\mid T}-\mat{P}_{t+1\mid t}\right)\mat{J}_t^{\top}.
\label{eq:ks_var}
\end{align}
The smoothed prevalence estimate is the first component of $\tilde{\alpha}_{t\mid T}$, namely $\tilde p_{t\mid T}$. 
Expanding the first component of \eqref{eq:ks_mean} gives
\[
\tilde p_{t\mid T}
=
\tilde p_{t\mid t}
+
[\mat{J}_t]_{11}\bigl(\tilde p_{t+1\mid T}-\tilde p_{t+1\mid t}\bigr)
+
[\mat{J}_t]_{12}\bigl(\tilde v_{t+1\mid T}-\tilde v_{t+1\mid t}\bigr).    
\]
Unlike the filtered estimate, $\tilde p_{t\mid T}$ combines information from both earlier and later observations and should therefore be interpreted as a retrospective summary rather than a real-time estimate.

It is also natural to compare the conditional variance of the latent prevalence level given the full observed series, $[\mat{P}_{t\mid T}]_{11}$, with the corresponding conditional variance given observations through time $t$, $[\mat{P}_{t\mid t}]_{11}$.
\begin{lem}
\label{lem:filter_smoother_variance}
When the filter and smoother use the same given $\mat{Q}$, the conditional variance of the latent prevalence level given the full observed series is no greater than the corresponding conditional variance given observations through time $t$:
\[
[\mat{P}_{t\mid T}]_{11}\le [\mat{P}_{t\mid t}]_{11}.
\]
\end{lem}
The proof is given in the Appendix.

\subsection{Missing Observations}
In some applications, daily prevalence estimates may be unavailable or may be treated as unreliable and therefore set to missing before filtering.
If either $\hat p_t$ or $R_t$ is unavailable, no measurement update is performed at time $t$, and the filter proceeds by prediction only:
\begin{equation}
\tilde{\alpha}_{t\mid t}=\tilde{\alpha}_{t\mid t-1},
\qquad
\mat{P}_{t\mid t}=\mat{P}_{t\mid t-1}.
\label{eq:kf_missing_update}
\end{equation}
Thus, missing days do not contribute an innovation term, but the level and slope states continue to evolve through the state equation until observations become available again.
This convention preserves continuity of the filtered prevalence trajectory without introducing ad-hoc interpolation.

For smoothing, no separate missing-data step is required. 
The backward recursions in \eqref{eq:ks_mean} and \eqref{eq:ks_var} are applied to the filtered and predicted quantities obtained from the forward Kalman filter recursion, regardless of whether an observation was available at time $t$. 
Thus, under a given $\mat{Q}$, a missing day receives a prediction based on past observations, whereas the smoothed estimate may also incorporate observations collected after that day through the backward recursion.

\subsection{Selection of the Process Variances}
\label{subsec:process-variance-selection}

The process-variance components are selected by maximizing the Gaussian innovation likelihood generated by the Kalman filter \citep{DeJong1988, Harvey1989, DurbinKoopman2012}.
For the filtered estimate at time $t$, define
\[
\mathcal{T}_{\mathrm{obs}}(t)
=
\{s:t_0<s\le t,\ \hat p_s\ \text{and}\ R_s\ \text{are observed}\}.
\]
For a candidate
$\mat{Q}=\operatorname{diag}(Q_{\text{level}},Q_{\text{slope}})$,
the forward recursion is run from $t_0$ through $t$ with $\mat{Q}$ held fixed.
This produces the innovations $\nu_s$ and innovation variances $S_s$ defined in \eqref{eq:kf_innovation} and \eqref{eq:kf_innovation_var}.
Their dependence on the candidate $\mat{Q}$ through the predicted state means and covariance matrices is left implicit.
Up to an additive constant, the Gaussian innovation log-likelihood based on all observations available through time $t$ is
\begin{equation}
\ell_t(Q_{\text{level}},Q_{\text{slope}})
=
-\frac{1}{2}
\sum_{s\in\mathcal{T}_{\mathrm{obs}}(t)}
\left\{
\log S_s
+
\frac{\nu_s^2}{S_s}
\right\}.
\label{eq:Q_innovation_ll}
\end{equation}
The process-variance estimates used by the KF at time $t$ are
\begin{align}
\left(
\widehat Q_{\text{level},t},
\widehat Q_{\text{slope},t}
\right)
&=
\arg\max_{Q_{\text{level}}\ge 0,\,Q_{\text{slope}}\ge 0}
\ell_t(Q_{\text{level}},Q_{\text{slope}}),
\label{eq:Q_mle}
\\
\widehat{\mat{Q}}_t
&=
\operatorname{diag}
\left(
\widehat Q_{\text{level},t},
\widehat Q_{\text{slope},t}
\right).
\notag
\end{align}
The filtered estimate $\tilde p_{t\mid t}$ is obtained by running the forward recursion through time $t$ with $\widehat{\mat{Q}}_t$ held fixed.
If fewer than two usable observations have accumulated, if either $\hat p_t$ or $R_t$ is unavailable, or if the optimization fails to converge, the most recent successfully estimated process-variance matrix is retained.
Before the first successful optimization, we use
$\widehat{\mat{Q}}_t=\operatorname{diag}(10^{-6},10^{-6})$.
The theoretical parameter space permits either or both variance components to be zero.
For numerical optimization, each component is restricted to $[10^{-10},10^{-2}]$, with an estimate at the lower bound treated as effectively zero.

Both $\widehat{\mat{Q}}_t$ and $\tilde p_{t\mid t}$ depend only on the HT estimates and observation variances available through time $t$.
The filtered estimate is therefore available in real time and does not use observations collected after $t$.
By contrast, the fixed-interval smoother estimates a single $\widehat{\mat{Q}}_T$ from the full observed series and uses it throughout the forward and backward recursions.
The smoothed estimate is therefore retrospective.

\subsection{Confidence Interval Constructions}
\label{subsec:ci-construction}
We begin with default 95\% confidence intervals centered at each point estimator and scaled by its corresponding standard error.
For the raw HT estimator, the standard error is based on the day-specific variance estimate $R_t$, whereas those for the joint KF and joint KS are obtained from the corresponding state covariance matrices.
Although these model-based intervals arise directly from the state-space formulation, the joint estimators may trade some bias for lower mean squared error, which can lead to undercoverage.
We therefore consider alternative confidence interval constructions.

\citet{kaplan2024confidence} study the construction of confidence intervals (CIs) in settings where a biased estimator is preferred because it may achieve a lower mean squared error than an unbiased alternative.
Their key observation is that, when an interval is centered at a biased estimator, its coverage depends not only on the estimator's standard error but also on the magnitude of its bias, the scale used to determine the interval width, and the choice of critical value.
This perspective is directly relevant here.
The raw HT estimator is approximately unbiased \citep{lee2026counterfactual}, whereas the joint KF and joint KS estimators may trade some bias for smoother trajectories and lower mean squared error and come with different uncertainty estimates.
We therefore consider several confidence interval constructions to examine how these different choices affect empirical coverage.

For the raw HT estimator, we use the same 95\% confidence interval as in \citet{lee2026counterfactual},
\begin{equation}
\hat p_t \pm 1.96\sqrt{R_t},
\label{eq:ht_ci}
\end{equation}
where $R_t$ is the day-specific delete-a-group jackknife variance estimate of $\hat p_t$.
For the joint filtered estimator, our default model-based 95\% confidence interval is
\begin{equation}
\tilde p_{t\mid t} \pm 1.96\sqrt{[\mat{P}_{t\mid t}]_{11}},
\label{eq:kf_ci_default}
\end{equation}
where $[\mat{P}_{t\mid t}]_{11}$ is the $(1,1)$ entry of the model-based filtered state covariance matrix.
For the joint smoothed estimator, the analogous model-based interval is
\begin{equation}
\tilde p_{t\mid T} \pm 1.96\sqrt{[\mat{P}_{t\mid T}]_{11}}.
\label{eq:ks_ci_default}
\end{equation}
For the KF, $\mat{P}_{t\mid t}$ is obtained by filtering through time $t$ using $\widehat{\mat{Q}}_t$.
For the KS, $\mat{P}_{t\mid T}$ is obtained using the full-series estimate $\widehat{\mat{Q}}_T$.
The covariance matrices and intervals treat the corresponding process-variance estimates as fixed and do not account for uncertainty from estimating them.

We also consider two alternative confidence-interval constructions for the KF and KS when the HT variance estimate $R_t$ is available on the target day.
Following the terminology of \citet{kaplan2024confidence}, we refer to these constructions as CI$_2$ and CI$_5$.
Under CI$_2$, the intervals are centered at the KF and KS point estimates while using the HT standard error:
\begin{equation}
\tilde p_{t\mid t} \pm 1.96\sqrt{R_t},
\qquad
\tilde p_{t\mid T} \pm 1.96\sqrt{R_t}.
\label{eq:ci2_style}
\end{equation}
Under CI$_5$, the intervals retain the same scale, $\sqrt{R_t}$, but replace the standard normal critical value with day-specific calibrated critical values:
\begin{equation}
\tilde p_{t\mid t} \pm \tilde z^{\mathrm{KF}}_{0.975,t}\sqrt{R_t},
\qquad
\tilde p_{t\mid T} \pm \tilde z^{\mathrm{KS}}_{0.975,t}\sqrt{R_t},
\label{eq:ci5_style}
\end{equation}
where $\tilde z^{\mathrm{KF}}_{0.975,t}$ and $\tilde z^{\mathrm{KS}}_{0.975,t}$ denote the corresponding day-specific CI$_5$ critical values.
Specifically, each day-specific critical value is chosen so that the corresponding interval attains the nominal coverage level when the HT estimator and the corresponding joint estimator have the same mean squared error.
The explicit formulas and computational procedure for obtaining these calibrated critical values are given in \citet{kaplan2024confidence}.
We apply their CI$_5$ construction separately to the joint KF and joint KS estimators.
For all $t$, these critical values satisfy $\tilde z^{\mathrm{KF}}_{0.975,t} \leq 1.96$ and $\tilde z^{\mathrm{KS}}_{0.975,t} \leq 1.96$, so the interval length under CI$_5$ is never greater than that under CI$_2$.
When either the HT estimate or its variance estimate $R_t$ is unavailable on day $t$, the CI$_2$ and CI$_5$ constructions are not directly applicable, whereas the default model-based KF and KS intervals remain well defined through the state-space covariance updates.

In addition to the default, CI$_2$, and CI$_5$ constructions for the joint estimators, we consider the CI$_6$ construction of \citet{kaplan2024confidence}, which changes both the interval calibration and the interval center through an optimal convex combination of two candidate estimators.
In our setting, this yields convex-combination estimators formed from the raw HT estimator and the corresponding joint estimator.
Specifically, the HT-KF mixture combines the raw HT estimator with the joint filtered estimate, and the HT-KS mixture combines the raw HT estimator with the joint smoothed estimate:
\begin{align}
\hat p^{\text{HT-KF}}_t
&=
(1-w^{\text{KF}}_t)\hat p_t + w^{\text{KF}}_t \tilde p_{t\mid t},
\label{eq:ht_kf_mixture}
\\
\hat p^{\text{HT-KS}}_t
&=
(1-w^{\text{KS}}_t)\hat p_t + w^{\text{KS}}_t \tilde p_{t\mid T},
\label{eq:ht_ks_mixture}
\end{align}
where $0 \le w^{\text{KF}}_t, w^{\text{KS}}_t \le 1$ are the corresponding day-specific CI$_6$ weights.
Under CI$_6$, the interval is centered at the mixture estimator and uses the HT standard error $\sqrt{R_t}$ as its scale.
To construct CI$_6$, we use the standard errors of the two estimators and their correlation to calibrate the critical value, and choose the weight to minimize interval length while maintaining nominal coverage. 
The explicit optimization criterion, critical-value calibration, and computational algorithm for the CI$_6$ weight are provided in \citet{kaplan2024confidence}.
We apply that procedure to the HT--KF and HT--KS pairs defined in \eqref{eq:ht_kf_mixture} and \eqref{eq:ht_ks_mixture}.
Although the filtered update in \eqref{eq:kf_update_prevalence_scalar} is also a convex combination of the one-step-ahead prediction $\tilde p_{t\mid t-1}$ and the raw HT estimate $\hat p_t$, we do not replace the Kalman gain $k_t$ by the CI$_6$ weight.
We explored such a replacement in preliminary analyses, but it did not yield satisfactory performance.
This is because $k_t$ is part of the filtering recursion, whereas the CI$_6$ weight is introduced only after the HT estimator and the corresponding KF or KS estimator have been computed.
We report the corresponding CI$_6$ mixture results in Section~\ref{subsec:simulation-results}.

For convenience, Table~\ref{tab:notation} summarizes the main notation used in the state-space formulation.
\begin{table}[h!]
\small
\centering
\caption{Notation used in the Kalman filtering and smoothing formulation.}
\label{tab:notation}
\begin{tabular}{lp{0.7\textwidth}}
\toprule
Symbol & Meaning \\
\midrule
$\hat p_t$ & Daily HT prevalence estimator at time $t$ \\
$\alpha_t = (p_t, v_t)^\top \in \mathbb{R}^2$ & Latent state vector at time $t$ \\
$p_t$ & Latent prevalence level at time $t$ \\
$v_t$ & Latent slope at time $t$ \\
$\vec{w}_t = (\eta_t, \zeta_t)^\top \in \mathbb{R}^2$ & State disturbance vector at time $t$ \\
$e_t$ & Observation error in the measurement equation \\
$\eta_t$ & Level disturbance \\
$\zeta_t$ & Slope disturbance \\
$\mat{I} \in \mathbb{R}^{2\times 2}$ & Identity matrix \\
$\mat{F} \in \mathbb{R}^{2\times 2}$ & State transition matrix \\
$\mat{H} \in \mathbb{R}^{1\times 2}$ & Observation matrix \\
$\mat{Q} \in \mathbb{R}^{2\times 2}$ & State disturbance covariance matrix \\
$R_t$ & Day-specific observation variance of $\hat p_t$, represented by its plug-in estimate in applications \\
$Q_{\text{level}}$ & Level-disturbance variance \\
$Q_{\text{slope}}$ & Slope-disturbance variance \\
$\widehat{\mat{Q}}_t \in \mathbb{R}^{2\times 2}$ & Process-variance matrix estimated using all observations available through time $t$ and used for the filtered estimate at time $t$ \\
$\tilde{\alpha}_{t\mid s} \in \mathbb{R}^2$ & Estimated state at time $t$ using observations through time $s$ \\
$\tilde p_{t\mid s}$ & Estimated prevalence level at time $t$ using observations through time $s$ \\
$\tilde v_{t\mid s}$ & Estimated slope at time $t$ using observations through time $s$ \\
$\mat{P}_{t\mid s} \in \mathbb{R}^{2\times 2}$ & State covariance matrix at time $t$ given observations through time $s$ \\
$\nu_t$ & Innovation at time $t$ \\
$S_t$ & Innovation variance at time $t$ \\
$\vec{K}_t \in \mathbb{R}^2$ & Kalman gain at time $t$ \\
$k_t$ & First component of $\vec{K}_t$ used in the prevalence update \\
$\mat{J}_t \in \mathbb{R}^{2\times 2}$ & Smoothing gain at time $t$ \\
$t_0$ & First time point at which both $\hat p_t$ and $R_t$ are observed \\
$T$ & Final time point in the analysis period \\
\bottomrule
\end{tabular}
\end{table}

\section{Simulation Studies}
\subsection{Design}
We conducted simulation studies to evaluate the behavior of the proposed filtering and smoothing procedures when applied to noisy daily prevalence estimates. 
The underlying disease and testing processes were adapted from the repeated testing framework of \citet{lee2026counterfactual}. 
That paper compared three daily prevalence summaries: the test-positive rate (TPR), an HT estimator that does not account for symptomatic testing and contact tracing, and an HT estimator that explicitly accounts for scheduled, symptomatic, and contact-tracing testing. 
Among these, only the last was approximately unbiased and closely tracked the true prevalence, which is the target quantity of interest. 
We therefore use that HT estimator here as the input series and study how the Kalman filter and corresponding smoother transform the resulting day-by-day estimates into more stable and interpretable prevalence trajectories.

A population of $10{,}000$ individuals with identically distributed processes was simulated over a 21-day period and organized into $5000$ exchangeable clusters of size 2, representing roommate-type living arrangements.
The hazard of initial exposure from outside the cluster for nonremoved individuals was modeled as
$h(\tau) = \frac{1}{10} \left( \frac{\tau(21-\tau)}{(21/2)^2} \left(\frac{1}{10}-\frac{1}{50}\right) + \frac{1}{50} \right),$
where $\tau$ denotes the time since day 0 or since the most recent clearance.
The within-cluster exposure hazard was set to $1/5$ times the number of infectious individuals in the same cluster and was assumed to act independently of external exposure.
The hazard for subsequent exposures was defined as $\frac{1}{2}h(\tau)$.
Simulations were initialized at 2\% prevalence and typically peaked near 5\%.
Test sensitivity and specificity were set to 83.2\% and 99.2\%, respectively, based on \citet{butler2021comparison}.
Following a positive test, individuals entered the Removed compartment for 5 days, after which they returned to the Well compartment.

We first describe scheduled testing governed by prespecified rules, followed by symptom-based and contact-tracing-based testing triggered by symptoms or confirmed exposures.
\begin{itemize}
  \item \textit{Simple random testing regimen:} nonremoved individuals are tested independently each day with probability $1/6$.
  \item \textit{Once-per-period regimen:} each nonremoved individual is eligible for one scheduled test per fixed-length calendar interval.
  \item \textit{Max-gap regimen:} the first scheduled test is uniformly distributed over the first 10 days, and subsequent scheduled testing probability increases quadratically with time since the most recent scheduled test or clearance.
  \item \textit{Min-max regimen:} similar to the max-gap regimen, but testing is prohibited within 5 days of the most recent scheduled test.
\end{itemize}
Exposed individuals became symptomatic on the first infectious day with probability $0.25$, whereas unexposed individuals could also trigger symptom-based testing with probability $0.01$.
Whenever an individual tested positive, all other nonremoved members of the same cluster were tested on the following day under the contact-tracing mechanism, regardless of scheduled testing eligibility.

We considered two scenarios: one using the complete sequence of daily HT estimates, and one in which testing was unavailable on days 10 and 11, so the HT estimator was treated as missing on those days and the filter proceeded by prediction only until observations resumed.
This second scenario was included to mimic short operational interruptions such as weekends with no testing, temporary laboratory delays, or brief administrative gaps in reporting.
For each scenario and each testing design, we generated 100 simulation replicates.
For each simulated dataset, the day-specific variance of the HT estimator was estimated using a delete-a-group jackknife procedure \citep{kott2001delete}, and the resulting variance estimates served as the observation variances $R_t$ in the joint state-space model.
At each day $t$, the joint KF used $\widehat{\mat{Q}}_t$ based on all available HT estimates and observation variances through that day and reported the corresponding filtered estimate.
The joint KS instead used the single process-variance estimate $\widehat{\mat{Q}}_T$ obtained from the full series.

We summarize performance using the mean estimated prevalence trajectory, the day-specific root mean squared error (RMSE), and the day-specific coverage of 95\% confidence intervals (CIs) for the raw HT estimator, the joint Kalman filter (KF), and the corresponding joint Kalman smoother (KS).
The interval constructions used throughout are defined in Section~\ref{subsec:ci-construction}.
For the joint estimators, these include the default model-based intervals in \eqref{eq:kf_ci_default}--\eqref{eq:ks_ci_default}, CI$_2$ in \eqref{eq:ci2_style}, and CI$_5$ in \eqref{eq:ci5_style}.
For the KF and KS, \textit{Coverage}, \textit{Coverage*}, and \textit{Coverage+} denote the empirical proportions of simulation replicates covered by the default model-based interval, CI$_2$, and CI$_5$, respectively.
Because CI$_2$ and CI$_5$ rely on the day-specific HT standard error, we report \textit{Coverage*} and \textit{Coverage+} only in the complete-data setting.

\subsection{Results}
\label{subsec:simulation-results}

\begin{figure}[!ht]
    \centering
    \includegraphics[width=1\textwidth]{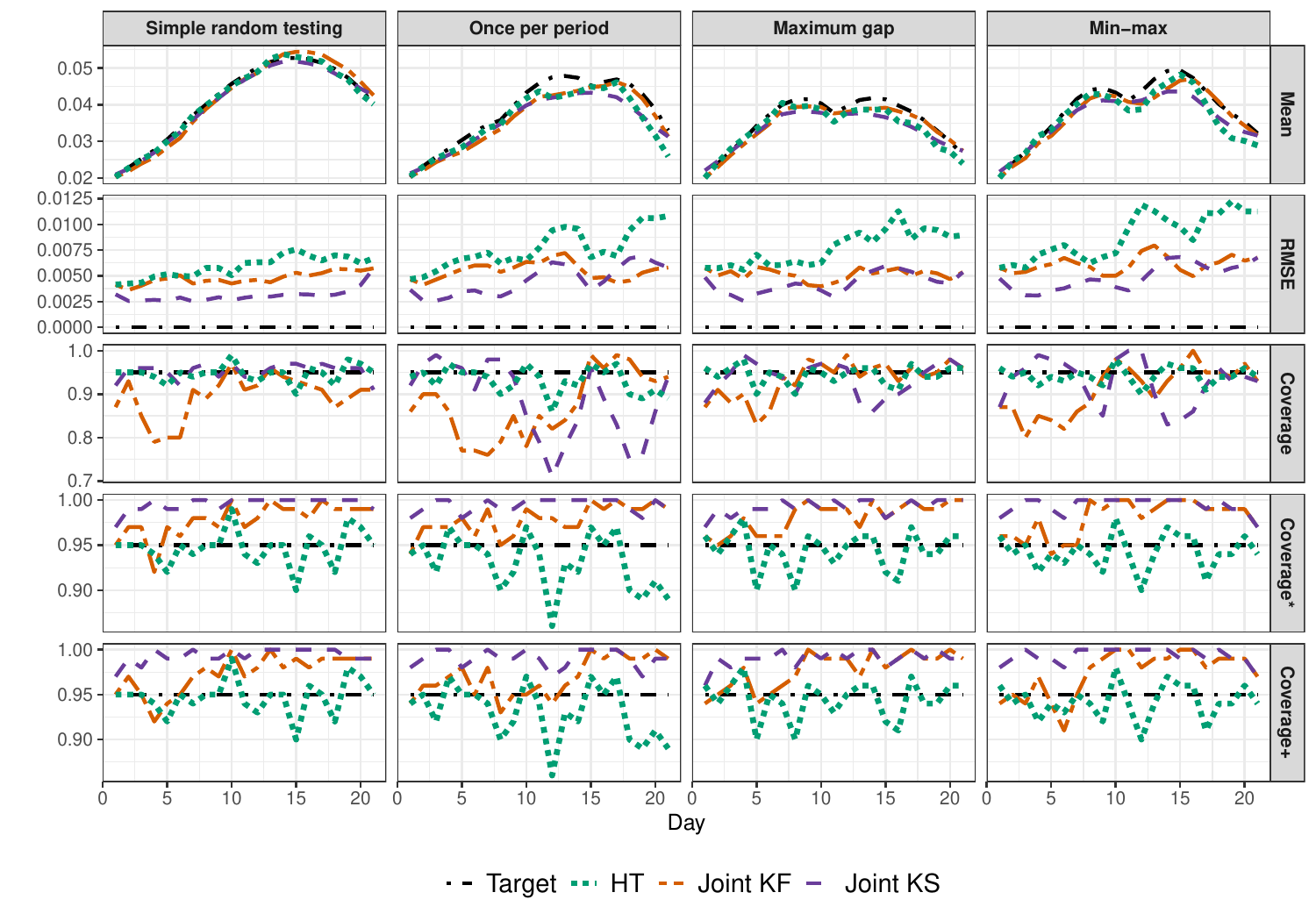}
    \caption{Simulations under the complete-data scenario for four scheduled testing designs with additional symptomatic and contact-tracing testing.
    For each design, the first row shows the mean estimated prevalence over 100 simulation replicates, the second row shows the corresponding day-specific RMSE relative to the true simulated prevalence, and the last three rows show the empirical coverage probabilities for the three interval constructions described in the text: Coverage (default model-based intervals), Coverage* (CI$_2$), and Coverage+ (CI$_5$).
    Curves represent the target prevalence (black, dotdash), the HT estimator (green, dotted), the real-time joint Kalman filter (KF) estimate using $\widehat{\mat{Q}}_t$ (orange, twodash), and the retrospective joint Kalman smoother (KS) estimate using $\widehat{\mat{Q}}_T$ (purple, dashed).}
    \label{fig:sim-prev-KS-KF}
\end{figure}

Figure~\ref{fig:sim-prev-KS-KF} compares the joint KF and KS under the complete-data scenario. 
The raw HT estimator tracks the overall pattern of the target prevalence, but its RMSE generally increases over time. 
As discussed in \citet{lee2026counterfactual}, the HT estimator exhibits a small downward bias in numerical implementation due to Jensen's inequality. 
Both joint methods preserve the main temporal pattern while substantially reducing RMSE relative to the raw HT estimator. 
Across days, the joint KS often attains the smallest RMSE, while the joint KF is typically slightly less accurate but still clearly improves on the raw HT estimator. 
The default model-based coverage results, shown in the third row labeled \textit{Coverage}, indicate that both the joint KF and joint KS exhibit undercoverage on some days.
The relative performance varies across days and testing designs, so neither method provides uniformly better coverage.
This pattern is consistent with \citet{kaplan2024confidence}, which shows that confidence intervals centered at a biased estimator and scaled by its own standard error, corresponding to CI$_4$ in that paper, can suffer from undercoverage when the bias is non-negligible relative to the estimator's variance.
The last two rows evaluate intervals that retain the KF and KS point estimates but use the HT standard error. 
Both \textit{Coverage*} and \textit{Coverage+}, corresponding to CI$_2$ and CI$_5$, are generally much higher than the default model-based coverage and are often close to 1, indicating substantial overcoverage and overly wide intervals.
Although \textit{Coverage+} uses a smaller critical value than \textit{Coverage*} and therefore produces shorter intervals, its empirical coverage remains close to 1 in many settings.

\begin{figure}[h!]
    \centering
    \includegraphics[width=1\textwidth]{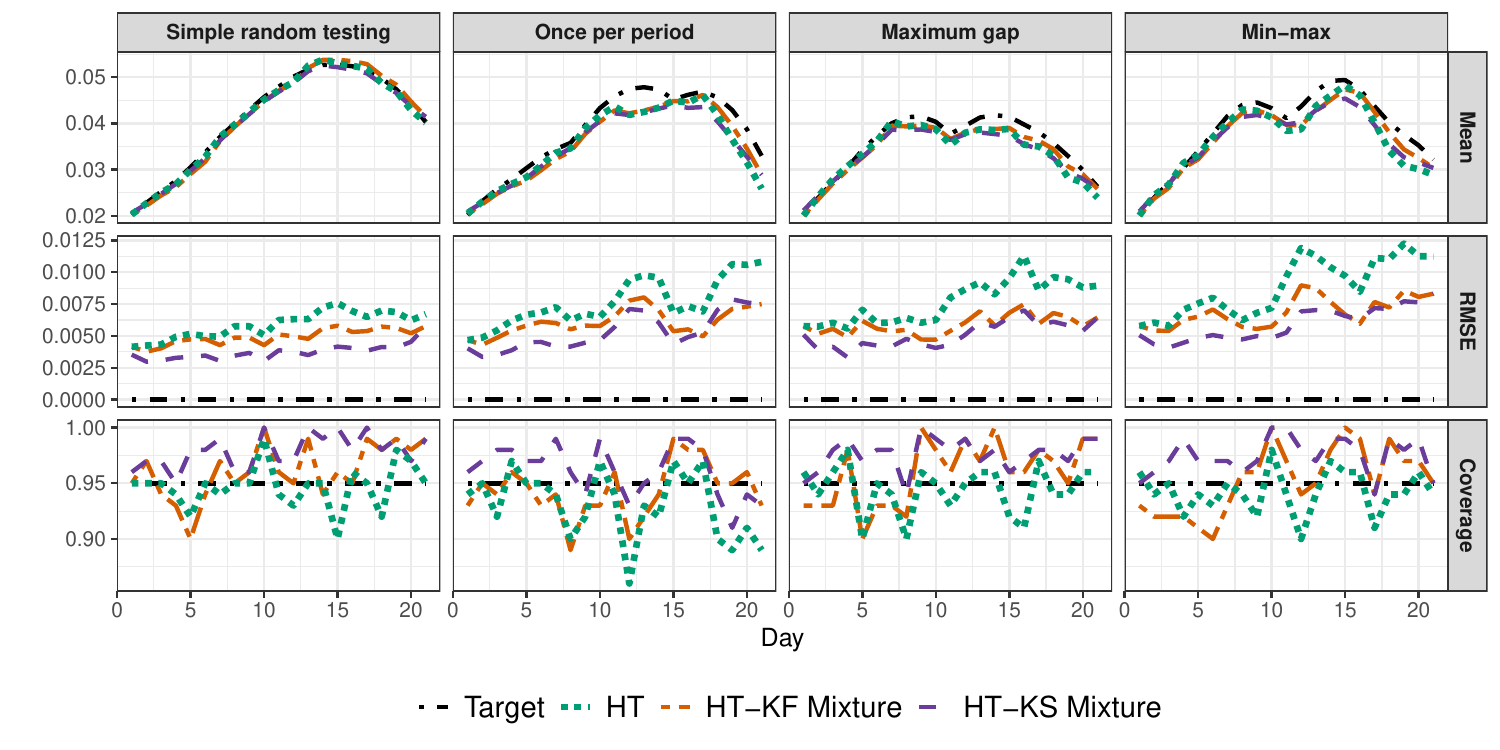}
    \caption{Complete-data comparison of the raw HT estimator with the HT-KF and HT-KS CI$_6$ convex-combination estimators across four testing designs and 100 simulation replicates.
    Rows show the mean estimated prevalence, day-specific RMSE, and nominal 95\% confidence-interval coverage.
    The HT-KF and HT-KS mixtures combine the HT estimator with the real-time joint KF based on $\widehat{\mat{Q}}_t$ and the retrospective joint KS based on $\widehat{\mat{Q}}_T$, respectively.}
    \label{fig:sim-mixture}
\end{figure}

We next consider the CI$_6$ convex-combination estimators defined in Section~\ref{subsec:ci-construction}. 
Figure~\ref{fig:sim-mixture} shows that the HT-KF and HT-KS mixtures preserve the main temporal pattern of the target prevalence while generally reducing RMSE relative to the raw HT estimator. 
The HT-KS mixture typically achieves a larger RMSE reduction than the HT-KF mixture, although both mixtures generally achieve smaller reductions than their corresponding joint estimators in Figure~\ref{fig:sim-prev-KS-KF}. 
Their empirical coverage is often closer to the nominal 95\% level than that of the joint KF and joint KS. 
Because CI$_6$ requires the HT variance estimate $R_t$, however, it is not directly available on days with missing HT observations.

\begin{figure}[h!]
    \centering
    \includegraphics[width=1\textwidth]{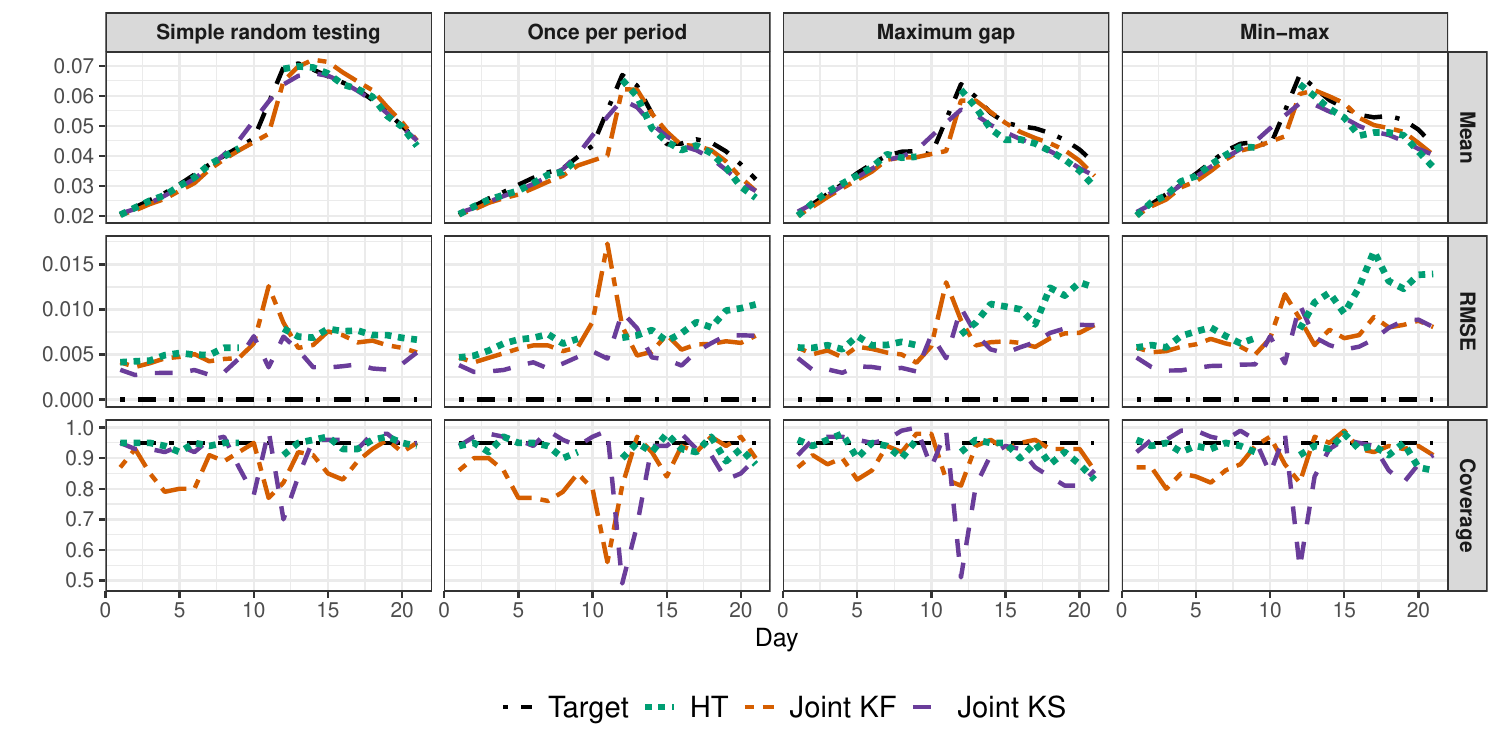}
    \caption{Simulations with testing unavailable on days 10 and 11. The layout and estimators are the same as in Figure~\ref{fig:sim-prev-KS-KF}, except that only the default model-based coverage row is shown.
    On these days, the HT estimator is treated as missing and its confidence interval is undefined.}
    \label{fig:sim-prev-missing}
\end{figure}

Figure~\ref{fig:sim-prev-missing} shows that a similar qualitative pattern holds when testing is unavailable on days 10 and 11. 
On those days, the raw HT estimator and its confidence interval are unavailable and therefore omitted, whereas both the joint KF and the joint KS continue to produce prevalence estimates. 
In this setting, we report the default model-based intervals for the joint methods, based on the corresponding filtered and smoothed state covariance matrices. 
During the missing interval, the joint KF retains the most recently estimated process variances and proceeds through prediction-only updates based on past observations, whereas the joint KS additionally uses later observations to reconstruct that interval retrospectively. 
Because testing is unavailable during this period, individuals who would otherwise have tested positive on days 10 and 11 are not detected and therefore do not enter isolation, which may lead to a higher underlying prevalence when testing resumes on day 12. 
Both joint estimators again reduce RMSE relative to the raw HT estimator, but the difference between them becomes more pronounced over the missing interval, where the joint KS often achieves substantially lower RMSE than the joint KF.
This improvement in point estimation is accompanied by poorer interval calibration, with the joint KS tending to exhibit the most severe undercoverage.

\begin{figure}[h!]
    \centering
    \includegraphics[width=1\textwidth]{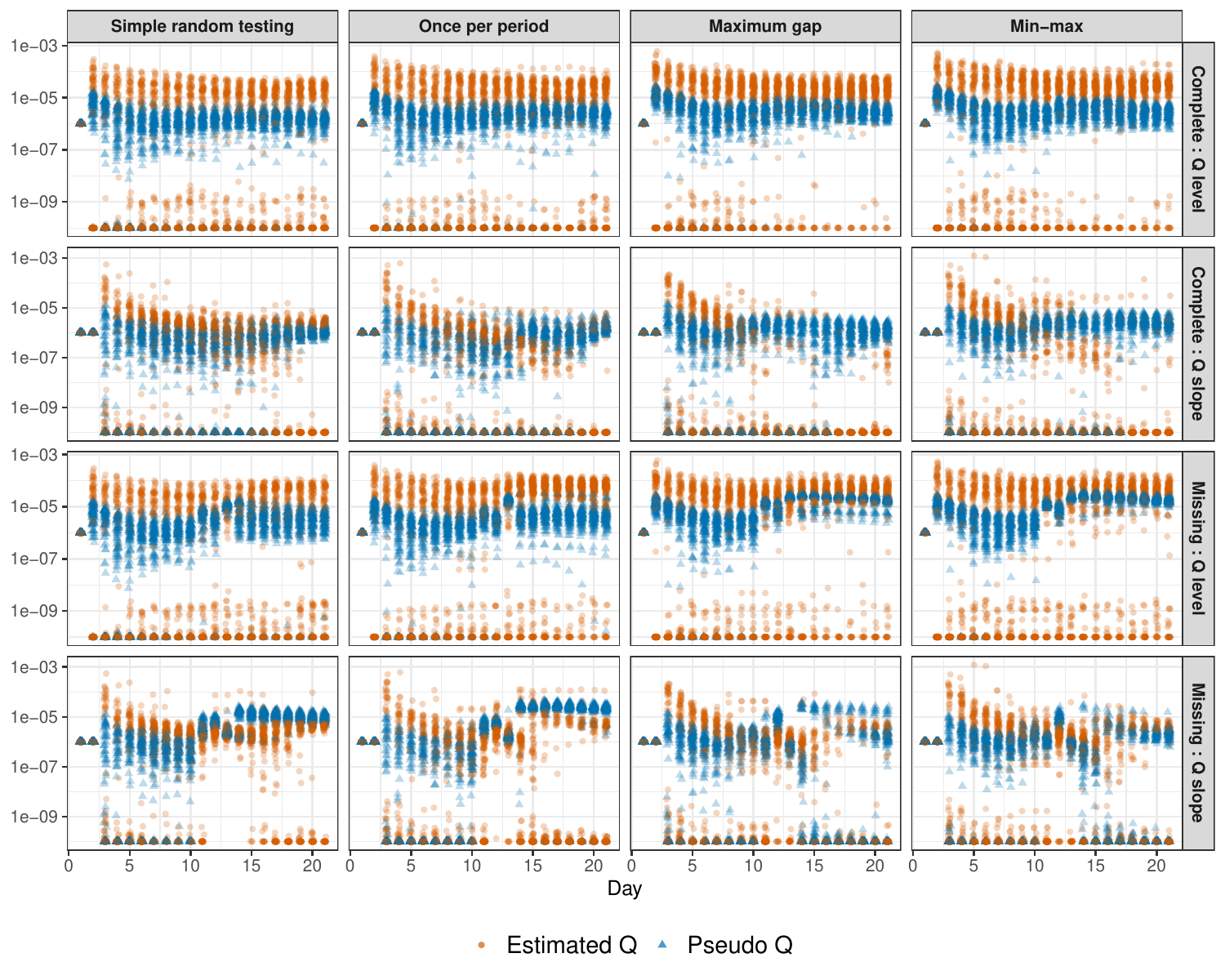}
    \caption{Day-specific process-variance estimates for the real-time joint KF, obtained using all observations available through each day (orange circles), and the corresponding pseudo-oracle estimates based on the true prevalence trajectory with negligible observation variance (blue triangles). 
    Columns identify the four testing designs, and rows identify the complete-data or missing-observation scenario and the level or slope component. 
    Each point represents one of 100 simulation replicates on the indicated day.
    The vertical scale is logarithmic, and estimates at $10^{-10}$ are treated as effectively zero.}
    \label{fig:q-history-comparison}
\end{figure}

Figure~\ref{fig:q-history-comparison} compares the day-specific process-variance estimates with pseudo-oracle values computed from the true prevalence trajectories. 
Let $p_t^\ast$ denote the true prevalence on day $t$ in a given simulation replicate. 
For each day $t$, we define the pseudo-oracle process variances as
\[
(\widehat Q_{\text{level},t}^\ast,\widehat Q_{\text{slope},t}^\ast)
=
\arg\max_{Q_{\text{level}}\ge 0,\,Q_{\text{slope}}\ge 0}
\ell_t^\ast(Q_{\text{level}},Q_{\text{slope}}),
\]
where $\ell_t^\ast$ is the Gaussian innovation log-likelihood obtained by fitting the same joint local linear trend model to $\{p_s^\ast\}_{s=1}^t$ with negligible observation variance. 
Most nonboundary estimates lie between $10^{-7}$ and $10^{-4}$, although some attain the numerical lower bound of $10^{-10}$. 
The pseudo-oracle estimates, shown by blue triangles, are generally more tightly concentrated across replicates, whereas the estimates based on the noisy HT series, shown by orange circles, exhibit greater dispersion. 
These patterns suggest that the process variances are best interpreted as data-adaptive tuning parameters that control the smoothness and local responsiveness of the filtered trajectory.

\section{Prevalence Estimates from OSU Fall 2020}
We used de-identified longitudinal testing data from 11,335 undergraduate students residing on campus during the Fall 2020 semester. 
Testing was performed through Student Health Services (SHS), Vault Health (VAULT), and the Applied Microbiology Services Laboratory (AMSL). 
Under the routine surveillance program, eligible students were required to complete one saliva PCR test during each Monday-Friday workweek, choosing a testing day within that interval through either VAULT or AMSL. 
SHS was used primarily for supplemental testing of symptomatic individuals and close contacts. 
Routine surveillance, symptomatic, and contact-tracing testing were generally not conducted on weekends or university holidays, and these days therefore appear as missing testing days in our analysis.
During the move-in period from August 14 to August 16, before the beginning of classes, all students were tested on arrival regardless of symptoms, and SHS and VAULT records were not distinguished. 
On average, symptomatic and contact-tracing tests made up about 1-2\% of daily tests, and daily testing volume was usually between 1,000 and 2,000. 
We assumed a 10-day clearance period after a positive test, during which individuals were excluded from testing, followed by an additional 80-day exemption from scheduled surveillance. 
Test sensitivity and specificity were set to 0.832 and 1, respectively, and the prevalence lower bound was truncated at zero.

\begin{figure}[h!]
    \centering
    \includegraphics[width=1\textwidth]{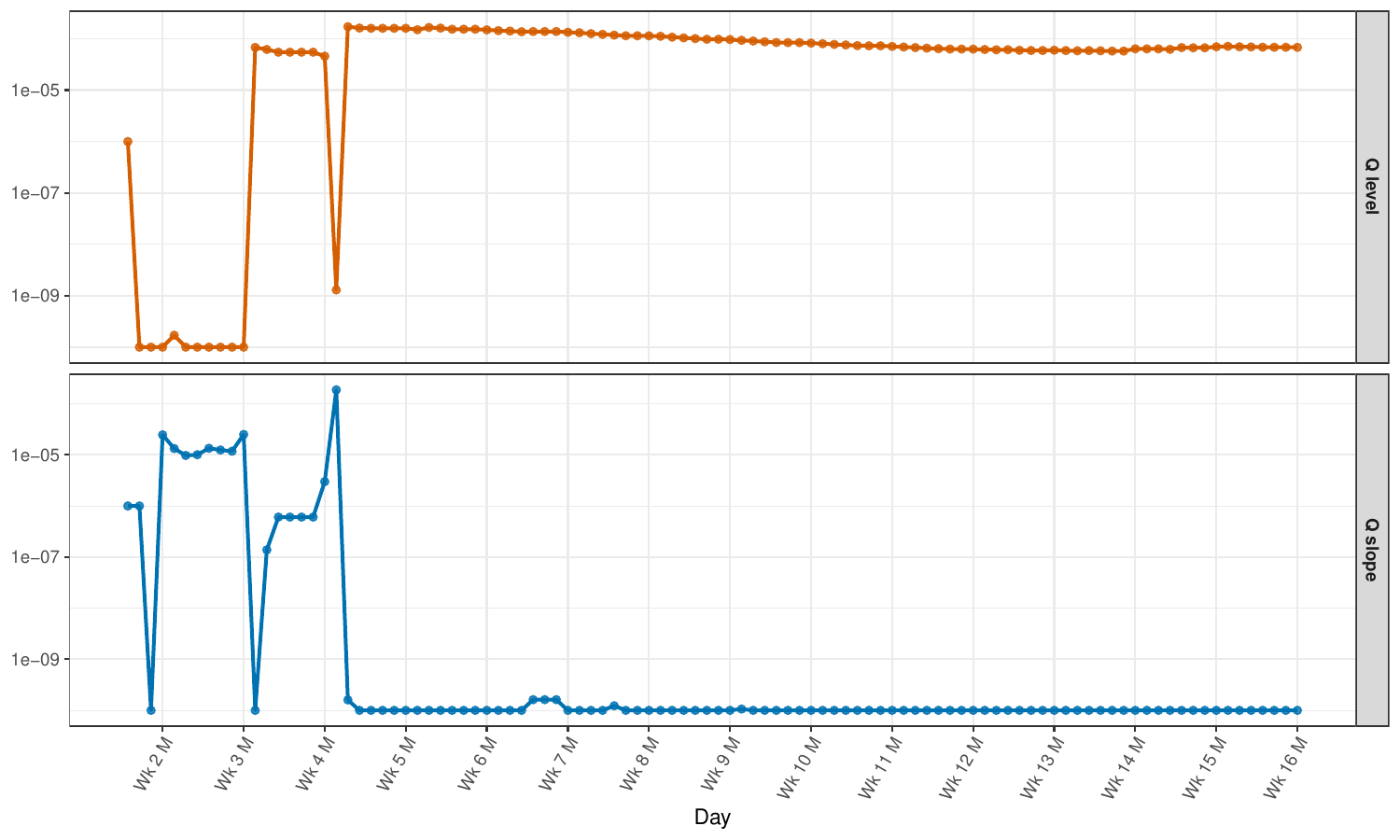}
    \caption{Process-variance values used by the real-time joint KF in the OSU analysis. 
    On each observed day for which optimization succeeds, $\widehat Q_{\text{level},t}$ and $\widehat Q_{\text{slope},t}$ are estimated using all HT estimates and observation variances available through that day.
    Otherwise, the starting or most recent successfully estimated values are retained. 
    The vertical scale is logarithmic, and values at $10^{-10}$ are treated as effectively zero.}  
    \label{fig:real-q-history}
\end{figure}

Figure~\ref{fig:real-q-history} shows how the process variances used by the joint KF change as information accumulates.
At day 102, the estimates are $\widehat Q_{\text{level},102}=6.81\times10^{-5}$ and $\widehat Q_{\text{slope},102}=10^{-10}$, with the latter treated as effectively zero.
The joint KS uses $\widehat{\mat{Q}}_T$, estimated from the full 102-day series, and is retrospective.

\begin{figure}[h!]
    \centering
    \includegraphics[width=1\textwidth]{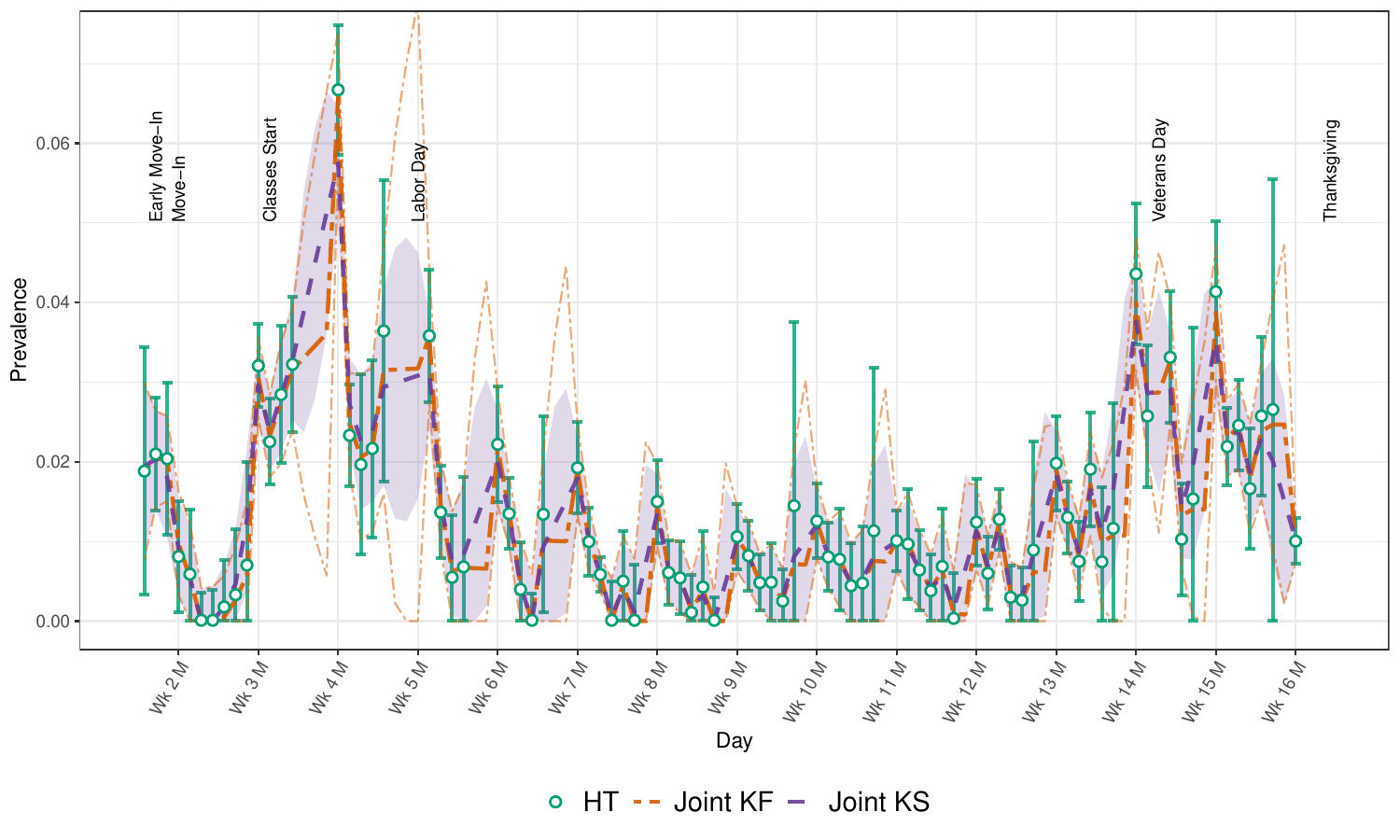}
    \caption{Daily prevalence estimates from the HT estimator, the real-time joint Kalman filter (KF), and the retrospective joint Kalman smoother (KS). 
    Points represent the HT estimator.
    The joint KF using $\widehat{\mat{Q}}_t$ estimated from all observations available through time $t$ and the joint KS using the full-series estimate $\widehat{\mat{Q}}_T$ are shown as twodash and dashed curves, respectively. 
    Green error bars show the 95\% confidence intervals for the HT estimator, the model-based 95\% confidence interval for the joint KF is shown by the pair of lighter twodash boundary curves, and the model-based 95\% confidence interval for the joint KS is shown by the shaded band.
    Key events, including early move-in, move-in, the start of classes, Labor Day, Veterans Day, and Thanksgiving, are annotated near the corresponding days at the top of the panel. 
    Days with fewer than 100 tests are treated as missing observations before filtering and smoothing, so the HT estimates are omitted on those days.}  
    \label{fig:real-prev-KF-KS}
\end{figure}

Figure~\ref{fig:real-prev-KF-KS} shows the HT, real-time joint KF, and retrospective joint KS prevalence trajectories for the OSU testing data. 
For most of the study period, the joint KF and joint KS estimates are nearly indistinguishable, suggesting that filtering and smoothing produce very similar prevalence trajectories when observations are available regularly. 
Relative to the raw HT estimator, both joint estimates are smoother and tend to lie lower around Friday of weeks 4, 6, 9, and 10, when the HT estimator shows sharper upward movement. 
A difference between the joint KF and joint KS is visible during periods with missing observations, particularly on weekends, Labor Day, and Veterans Day, when no testing was conducted and the HT estimator is therefore unavailable. 
Their model-based 95\% confidence intervals are also similar on most observed days, but the difference becomes more apparent over weekends and holidays, when the KS interval is generally narrower than the KF interval. 
Prevalence estimates on Mondays are often higher than those on the preceding Fridays. 
This may reflect the absence of testing on Saturdays and Sundays, so that infections accumulating over the weekend are not detected until testing resumes at the start of the following week, a pattern broadly consistent with the simulation results under missing observations. 
Alternatively, it may suggest that transmission occurs during the weekend rather than primarily during class days.

\section{Discussion}
We used a joint local linear trend model to obtain more stable prevalence trajectories from noisy daily HT estimates while incorporating day-specific observation variances and accommodating missing estimates. 
In the simulations, both the joint KF and joint KS reduced RMSE, with the KS often attaining the smallest values. 
Both methods exhibited undercoverage on some days, and neither provided uniformly better coverage. 
CI$_2$ and CI$_5$ generally produced overcoverage, whereas the CI$_6$ mixtures often achieved coverage closer to the nominal level but smaller RMSE reductions than the corresponding joint estimators. 
In the OSU application, the joint KF and joint KS produced similar trajectories on most observed days, with clearer differences during periods without testing.

At time $t$, the joint KF uses $\widehat{\mat{Q}}_t$, determined solely from the HT estimates and observation variances available through that time.
The resulting filtered prevalence estimate uses no observations collected after time $t$ and is therefore available in real time.
By contrast, the joint KS estimates $\widehat{\mat{Q}}_T$ from the full series and uses later observations, making it retrospective.
One-sided moving averages can also be updated sequentially, whereas centered moving averages and LOESS are retrospective.
Unlike the joint KF and KS, these methods do not directly incorporate day-specific observation variances or the state-space prediction mechanism for missing observations.

\section*{Reproducibility and Code Availability}
All analyses were implemented using the \texttt{PrevKalman} R package, which we developed to organize the Horvitz--Thompson prevalence estimator, block jackknife confidence intervals, and Kalman filtering and smoothing procedures. 
The source code for the package is available at \url{https://github.com/Jeongjin95/PrevKalman}.

\section*{Funding}
This research received no external funding.

\section*{Conflicts of Interest}
The authors report no conflicts of interest.

\bibliographystyle{chicago}
\bibliography{ref}

@article{Kalman1960,
  author = {Kalman, R. E.},
  title = {A New Approach to Linear Filtering and Prediction Problems},
  journal = {Journal of Basic Engineering},
  volume = {82},
  number = {1},
  pages = {35--45},
  year = {1960},
  doi = {10.1115/1.3662552}
}

@article{DeJong1988,
  author = {De Jong, Piet},
  title = {The Likelihood for a State Space Model},
  journal = {Biometrika},
  volume = {75},
  number = {1},
  pages = {165--169},
  year = {1988}
}

@book{Harvey1989,
  author = {Harvey, Andrew C.},
  title = {Forecasting, Structural Time Series Models and the Kalman Filter},
  publisher = {Cambridge University Press},
  address = {Cambridge},
  year = {1989}
}

@book{DurbinKoopman2012,
  author = {Durbin, James and Koopman, Siem Jan},
  title = {Time Series Analysis by State Space Methods},
  edition = {2},
  publisher = {Oxford University Press},
  address = {Oxford},
  year = {2012}
}

@article{baker2022successful,
  author = {Baker, Michael G and Wilson, Nick and Anglemyer, Andrew},
  title = {Successful elimination of {COVID}-19 transmission in {N}ew {Z}ealand},
  journal = {New England Journal of Medicine},
  volume = {383},
  number = {8},
  pages = {e56},
  year = {2020}
}

@article{butler2021comparison,
  title = {Comparison of saliva and nasopharyngeal swab nucleic acid amplification testing for detection of {SARS-CoV-2}: a systematic review and meta-analysis},
  author = {Butler-Laporte, Guillaume and Lawandi, Alexander and Schiller, Ian and Yao, Mandy and Dendukuri, Nandini and McDonald, Emily G and Lee, Todd C},
  journal = {JAMA Internal Medicine},
  volume = {181},
  number = {3},
  pages = {353--360},
  year = {2021}
}

@article{chang2021repeat,
  title = {Repeat {SARS-CoV-2} testing models for residential college populations},
  author = {Chang, Joseph T and Crawford, Forrest W and Kaplan, Edward H},
  journal = {Health Care Management Science},
  volume = {24},
  pages = {305--318},
  year = {2021}
}

@article{college21,
  author = {Paltiel, A David and Schwartz, Jason L},
  title = {Assessing {COVID-19} prevention strategies to permit the safe opening of residential colleges in fall 2021},
  journal = {Annals of Internal Medicine},
  volume = {174},
  number = {11},
  pages = {1563--1571},
  year = {2021}
}

@article{efron1987better,
  title = {Better bootstrap confidence intervals},
  author = {Efron, Bradley},
  journal = {Journal of the American Statistical Association},
  volume = {82},
  number = {397},
  pages = {171--185},
  year = {1987}
}

@article{kott2001delete,
  title = {The delete-a-group jackknife},
  author = {Kott, Phillip S},
  journal = {Journal of Official Statistics},
  volume = {17},
  number = {4},
  pages = {521--526},
  year = {2001}
}

@article{mba21,
  author = {Mack, Christina D and DiFiori, John and Tai, Caroline G and Shiue, Kristin Y and Grad, Yonatan H and Anderson, Deverick J and Ho, David D and Sims, Leroy and LeMay, Christopher and Mancell, Jimmie and others},
  title = {{SARS-CoV-2} transmission risk among {N}ational {B}asketball {A}ssociation players, staff, and vendors exposed to individuals with positive test results after {COVID-19} recovery during the 2020 regular and postseason},
  journal = {JAMA Internal Medicine},
  volume = {181},
  number = {7},
  pages = {960--966},
  year = {2021}
}

@article{school21,
  author = {Schultes, Olivia and Clarke, Victoria and Paltiel, A David and Cartter, Matthew and Sosa, Lynn and Crawford, Forrest W},
  title = {{COVID-19} Testing and Case Rates and Social Contact Among Residential College Students in {C}onnecticut During the 2020--2021 Academic Year},
  journal = {JAMA Network Open},
  volume = {4},
  number = {12},
  pages = {e2140602},
  year = {2021}
}

@article{lee2026counterfactual,
author  = {Lee, Jeongjin and Yang, Junke and Rempala, Grzegorz A. and Schnell, Patrick M.},
title   = {A Counterfactual Framework for Estimating Infectious Disease Prevalence under Repeated Testing with Symptomatic and Contact-Tracing Components},
journal = {Annals of Applied Statistics},
note    = {Accepted for publication},
year    = {2026}
}

@article{schnell2024overcoming,
  title = {Overcoming repeated testing schedule bias in estimates of disease prevalence},
  author = {Schnell, Patrick M and Wascher, Matthew and Rempala, Grzegorz A},
  journal = {Journal of the American Statistical Association},
  volume = {119},
  number = {545},
  pages = {1--13},
  year = {2024}
}

@article{work22,
  author = {Rosella, Laura C and Agrawal, Ajay and Gans, Joshua and Goldfarb, Avi and Sennik, Sonia and Stein, Janice},
  title = {Large-scale implementation of rapid antigen testing system for {COVID-19} in workplaces},
  journal = {Science Advances},
  volume = {8},
  number = {8},
  pages = {eabm3608},
  year = {2022}
}

@book{hernan2020causal,
  author={Hernán, Miguel A. and Robins, James M.},
  title={Causal Inference: What If},
  year={2020},
  publisher={Chapman \& Hall/CRC}
}

@article{horvitz1952generalization,
  title={A generalization of sampling without replacement from a finite universe},
  author={Horvitz, Daniel G and Thompson, Donovan J},
  journal={Journal of the American Statistical Association},
  volume={47},
  number={260},
  pages={663--685},
  year={1952},
  publisher={Taylor \& Francis}
}

@article{kaplan2024confidence,
  title={Confidence intervals for intentionally biased estimators},
  author={Kaplan, David M and Liu, Xin},
  journal={Econometric Reviews},
  volume={43},
  number={2-4},
  pages={197--214},
  year={2024},
  publisher={Taylor \& Francis}
}

\appendix

\section{Derivation}

Throughout the appendix, the filter starts at $t_0$ defined in \eqref{eq:kf_t0}, uses prediction-only steps for missing observations, and conditions only on observations available by each time point.
We treat $Q_{\text{level}}$, $Q_{\text{slope}}$, the observation variances, and the missingness pattern as fixed, and assume that $\{\vec{w}_t\}$ and $\{e_t\}$ are independent over time, mutually independent, and independent of the initial state.
Thus, the derivations in this section apply to any given $\mat{Q}$. 
For the real-time KF through time $u$, they are applied to the observations available through $u$ with $\mat{Q}=\widehat{\mat{Q}}_u$ treated as fixed.
For the KS, they are applied to the full series with $\mat{Q}=\widehat{\mat{Q}}_T$.

\subsection{Derivation of the Kalman Filter Recursion}

Assume that for some $t\ge t_0+1$,
\begin{equation}
\alpha_{t-1}\mid \hat p_1,\dots,\hat p_{t-1}
\sim
N\left(\tilde{\alpha}_{t-1\mid t-1},\mat{P}_{t-1\mid t-1}\right).
\label{eq:app_prev_filter}
\end{equation}
We derive the prediction and update steps from \eqref{eq:kf_state} and \eqref{eq:kf_meas}.

\paragraph{Prediction step.}
By the state equation,
\[
\alpha_t=\mat{F}\alpha_{t-1}+\vec{w}_t,
\qquad
\vec{w}_t\sim N(\vec{0},\mat{Q}),
\]
where $\vec{w}_t$ is independent of $\hat p_1,\dots,\hat p_{t-1}$ and of $\alpha_{t-1}$.
Taking conditional expectations given $\hat p_1,\dots,\hat p_{t-1}$ yields
\begin{align}
\tilde{\alpha}_{t\mid t-1}
&=
\E[\alpha_t\mid \hat p_1,\dots,\hat p_{t-1}]
\notag
\\
&=
\E[\mat{F}\alpha_{t-1}+\vec{w}_t\mid \hat p_1,\dots,\hat p_{t-1}]
\notag
\\
&=
\mat{F}\E[\alpha_{t-1}\mid \hat p_1,\dots,\hat p_{t-1}]
+
\E[\vec{w}_t]
\notag
\\
&=
\mat{F}\tilde{\alpha}_{t-1\mid t-1},
\end{align}
because $\E[\vec{w}_t]=\vec{0}$.
Similarly, using independence of $\vec{w}_t$ and $\alpha_{t-1}$ conditional on the past,
\begin{align}
\mat{P}_{t\mid t-1}
&=
\Var[\alpha_t\mid \hat p_1,\dots,\hat p_{t-1}]
\notag
\\
&=
\Var[\mat{F}\alpha_{t-1}+\vec{w}_t\mid \hat p_1,\dots,\hat p_{t-1}]
\notag
\\
&=
\mat{F}\Var[\alpha_{t-1}\mid \hat p_1,\dots,\hat p_{t-1}]\mat{F}^{\top}
+
\Var[\vec{w}_t]
\notag
\\
&=
\mat{F}\mat{P}_{t-1\mid t-1}\mat{F}^{\top}+\mat{Q}.
\end{align}
Hence the predictive distribution is
\begin{equation}
\alpha_t\mid \hat p_1,\dots,\hat p_{t-1}
\sim
N\left(\tilde{\alpha}_{t\mid t-1},\mat{P}_{t\mid t-1}\right).
\label{eq:app_predictive_state}
\end{equation}

\paragraph{Measurement equation and innovation.}
Given $\alpha_t$, the observation equation implies
\begin{equation}
\hat p_t\mid \alpha_t,\hat p_1,\dots,\hat p_{t-1}
\sim
N\left(\mat{H}\alpha_t,R_t\right).
\label{eq:app_measurement}
\end{equation}
Combining \eqref{eq:app_predictive_state} and \eqref{eq:app_measurement}, the conditional mean of $\hat p_t$ given the past is
\begin{align}
\E[\hat p_t\mid \hat p_1,\dots,\hat p_{t-1}]
&=
\E[\mat{H}\alpha_t+e_t\mid \hat p_1,\dots,\hat p_{t-1}]
\notag
\\
&=
\mat{H}\tilde{\alpha}_{t\mid t-1},
\end{align}
because $\E[e_t]=0$.
Its conditional variance is
\begin{align}
\Var[\hat p_t\mid \hat p_1,\dots,\hat p_{t-1}]
&=
\Var[\mat{H}\alpha_t+e_t\mid \hat p_1,\dots,\hat p_{t-1}]
\notag
\\
&=
\mat{H}\mat{P}_{t\mid t-1}\mat{H}^{\top}+R_t
\notag
\\
&=
S_t.
\end{align}
Therefore,
\begin{equation}
\hat p_t\mid \hat p_1,\dots,\hat p_{t-1}
\sim
N\left(\mat{H}\tilde{\alpha}_{t\mid t-1},S_t\right),
\label{eq:app_predictive_obs}
\end{equation}
and the innovation is
\[
\nu_t
=
\hat p_t-\mat{H}\tilde{\alpha}_{t\mid t-1}.
\]
\paragraph{Update step.}
The update step corrects the predicted state using the innovation $\nu_t$.
Consider the joint Gaussian vector
\[
\begin{pmatrix}
\alpha_t\\
\hat p_t
\end{pmatrix}
\Bigg| \hat p_1,\dots,\hat p_{t-1}.
\]
From \eqref{eq:app_predictive_state} and \eqref{eq:app_predictive_obs}, it has conditional mean
\[
\begin{pmatrix}
\tilde{\alpha}_{t\mid t-1}\\
\mat{H}\tilde{\alpha}_{t\mid t-1}
\end{pmatrix}
\]
and block covariance matrix
\[
\begin{pmatrix}
\Var[\alpha_t\mid \hat p_1,\dots,\hat p_{t-1}] &
\Cov\{\alpha_t,\hat p_t\mid \hat p_1,\dots,\hat p_{t-1}\} \\
\Cov\{\hat p_t,\alpha_t\mid \hat p_1,\dots,\hat p_{t-1}\} &
\Var[\hat p_t\mid \hat p_1,\dots,\hat p_{t-1}]
\end{pmatrix}.
\]
For a jointly Gaussian vector
\[
\begin{pmatrix}
A\\
B
\end{pmatrix}
\sim
N\left(
\begin{pmatrix}
\mu_A\\
\mu_B
\end{pmatrix},
\begin{pmatrix}
\Sigma_{AA} & \Sigma_{AB}\\
\Sigma_{BA} & \Sigma_{BB}
\end{pmatrix}
\right),
\]
the conditional mean and conditional covariance of $A$ given $B$ are
\begin{align}
\E[A\mid B]
&=
\mu_A+\Sigma_{AB}\Sigma_{BB}^{-1}(B-\mu_B),
\label{eq:app_cond_gaussian_mean}
\\
\Var[A\mid B]
&=
\Sigma_{AA}-\Sigma_{AB}\Sigma_{BB}^{-1}\Sigma_{BA}.
\label{eq:app_cond_gaussian_var}
\end{align}
In our setting, $A$ is the latent state $\alpha_t$ and $B$ is the new observation $\hat p_t$.
Applying \eqref{eq:app_cond_gaussian_mean} with
\[
A=\alpha_t,
\qquad
B=\hat p_t,
\qquad
\mu_A=\tilde{\alpha}_{t\mid t-1},
\qquad
\mu_B=\mat{H}\tilde{\alpha}_{t\mid t-1},
\]
gives
\begin{equation}
\tilde{\alpha}_{t\mid t}
=
\tilde{\alpha}_{t\mid t-1}
+
\Cov\{\alpha_t,\hat p_t\mid \hat p_1,\dots,\hat p_{t-1}\}
\Var[\hat p_t\mid \hat p_1,\dots,\hat p_{t-1}]^{-1}
\nu_t.
\label{eq:app_gaussian_update}
\end{equation}
Specifically,
\begin{align}
& \Cov\{\alpha_t,\hat p_t\mid \hat p_1,\dots,\hat p_{t-1}\} \\
&=
\Cov\{\alpha_t,\mat{H}\alpha_t+e_t\mid \hat p_1,\dots,\hat p_{t-1}\}
\notag
\\
&=
\Cov\{\alpha_t,\mat{H}\alpha_t\mid \hat p_1,\dots,\hat p_{t-1}\}
\notag
\\
&=
\mat{P}_{t\mid t-1}\mat{H}^{\top},
\end{align}
since $e_t$ is independent of $\alpha_t$ and the past.
Therefore, the weight in \eqref{eq:app_gaussian_update} is
\[
\mat{P}_{t\mid t-1}\mat{H}^{\top}\Var[\hat p_t\mid \hat p_1,\dots,\hat p_{t-1}]^{-1}
=
\mat{P}_{t\mid t-1}\mat{H}^{\top}S_t^{-1}.
\]
Substituting into \eqref{eq:app_gaussian_update} gives
\[
\tilde{\alpha}_{t\mid t}
=
\tilde{\alpha}_{t\mid t-1}
+
\mat{P}_{t\mid t-1}\mat{H}^{\top}S_t^{-1}\nu_t.
\]
Defining
\[
\vec{K}_t=\mat{P}_{t\mid t-1}\mat{H}^{\top}S_t^{-1},
\]
we obtain
\[
\tilde{\alpha}_{t\mid t}
=
\tilde{\alpha}_{t\mid t-1}
+
\vec{K}_t\nu_t.
\]
Thus the updated state equals the predicted state plus the innovation multiplied by the Kalman gain.

For the covariance update, apply \eqref{eq:app_cond_gaussian_var} with $A=\alpha_t$ and $B=\hat p_t$ to obtain
\begin{align}
\mat{P}_{t\mid t}
&=
\mat{P}_{t\mid t-1}
-
\Cov\{\alpha_t,\hat p_t\mid \hat p_1,\dots,\hat p_{t-1}\}
\\
&\quad\times
\Var[\hat p_t\mid \hat p_1,\dots,\hat p_{t-1}]^{-1}
\\
&\quad\times
\Cov\{\hat p_t,\alpha_t\mid \hat p_1,\dots,\hat p_{t-1}\}
\notag
\\
&=
\mat{P}_{t\mid t-1}
-
\mat{P}_{t\mid t-1}\mat{H}^{\top}S_t^{-1}\mat{H}\mat{P}_{t\mid t-1}
\notag
\\
&=
\left(\mat{I}-\vec{K}_t\mat{H}\right)\mat{P}_{t\mid t-1}.
\end{align}
The subtraction term is positive semidefinite, so the update reduces uncertainty after observing $\hat p_t$.
The matrix $\mat{P}_{t\mid t-1}$ describes uncertainty before the update, whereas $\mat{P}_{t\mid t}$ describes uncertainty after incorporating the new information.
This establishes the Kalman recursion for the joint local linear trend model.

\subsection{Proof of Lemma~\ref{lem:ht_filter_variance}}

From the measurement equation, the day-specific observation variance of the raw HT estimator is $R_t$.
Because $\mat{H}= \begin{pmatrix} 1 & 0 \end{pmatrix},$ the innovation variance is
\[
S_t
=
\mat{H}\mat{P}_{t\mid t-1}\mat{H}^{\top}+R_t
=
[\mat{P}_{t\mid t-1}]_{11}+R_t.
\]
Since $\mat{P}_{t\mid t-1}$ is a covariance matrix, it is positive semidefinite, so $[\mat{P}_{t\mid t-1}]_{11}\ge 0$. 
Together with $R_t>0$, this implies that $S_t>0$.

Using the covariance update formula \eqref{eq:kf_update_var}, the $(1,1)$ entry of the filtered covariance matrix is
\begin{equation}
[\mat{P}_{t\mid t}]_{11}
=
[\mat{P}_{t\mid t-1}]_{11}
-
\frac{[\mat{P}_{t\mid t-1}]_{11}^2}{[\mat{P}_{t\mid t-1}]_{11}+R_t}
=
\frac{[\mat{P}_{t\mid t-1}]_{11}R_t}{[\mat{P}_{t\mid t-1}]_{11}+R_t}.
\label{eq:app_ht_filter_variance}
\end{equation}
Because $[\mat{P}_{t\mid t-1}]_{11}\ge 0$ and $R_t>0$, we have
\[
0
\le
\frac{[\mat{P}_{t\mid t-1}]_{11}}{[\mat{P}_{t\mid t-1}]_{11}+R_t}
\le 1.
\]
Multiplying by $R_t$ gives
\[
[\mat{P}_{t\mid t}]_{11}
=
R_t
\frac{[\mat{P}_{t\mid t-1}]_{11}}{[\mat{P}_{t\mid t-1}]_{11}+R_t}
\le R_t.
\]
This proves Lemma~\ref{lem:ht_filter_variance}.

\subsection{Derivation of the Kalman Smoother Recursion}

The smoother derivation uses the Markov property that, given $\alpha_{t+1}$ and the observations available through time $t$, future observations provide no additional information about $\alpha_t$, which justifies the conditional-moment decompositions used to obtain \eqref{eq:ks_mean} and \eqref{eq:ks_var}.

\paragraph{Conditional distribution of $\alpha_t$ given $\alpha_{t+1}$.}
Conditional on $\hat p_1,\dots,\hat p_t$, the state equation implies
\[
\alpha_{t+1}=\mat{F}\alpha_t+\vec{w}_{t+1},
\qquad
\vec{w}_{t+1}\sim N(\vec{0},\mat{Q}),
\]
with $\vec{w}_{t+1}$ independent of $\alpha_t$ and of the observed history through time $t$.
Therefore, the joint conditional distribution of $(\alpha_t,\alpha_{t+1})$ given $\hat p_1,\dots,\hat p_t$ is Gaussian:
\begin{equation}
\begin{pmatrix}
\alpha_t\\
\alpha_{t+1}
\end{pmatrix}
\Bigg| \hat p_1,\dots,\hat p_t
\sim
N\left(
\begin{pmatrix}
\tilde{\alpha}_{t\mid t}\\
\tilde{\alpha}_{t+1\mid t}
\end{pmatrix},
\begin{pmatrix}
\mat{P}_{t\mid t} & \mat{P}_{t\mid t}\mat{F}^{\top}\\
\mat{F}\mat{P}_{t\mid t} & \mat{P}_{t+1\mid t}
\end{pmatrix}
\right).
\label{eq:app_smoother_joint}
\end{equation}
Here the off-diagonal block is
\[
\Cov(\alpha_t,\alpha_{t+1}\mid \hat p_1,\dots,\hat p_t)
=
\Cov\{\alpha_t,\mat{F}\alpha_t+\vec{w}_{t+1}\mid \hat p_1,\dots,\hat p_t\}
=
\mat{P}_{t\mid t}\mat{F}^{\top}.
\]
Applying \eqref{eq:app_cond_gaussian_mean}--\eqref{eq:app_cond_gaussian_var} to \eqref{eq:app_smoother_joint}, with
\[
\mu_A=\tilde{\alpha}_{t\mid t},
\quad
\mu_B=\tilde{\alpha}_{t+1\mid t},
\quad
\Sigma_{AA}=\mat{P}_{t\mid t},
\quad
\Sigma_{AB}=\mat{P}_{t\mid t}\mat{F}^{\top},
\quad
\Sigma_{BB}=\mat{P}_{t+1\mid t},
\]
yields
\begin{align}
\E[\alpha_t\mid \alpha_{t+1},\hat p_1,\dots,\hat p_t]
&=
\tilde{\alpha}_{t\mid t}
+
\mat{P}_{t\mid t}\mat{F}^{\top}\mat{P}_{t+1\mid t}^{-1}
\left(\alpha_{t+1}-\tilde{\alpha}_{t+1\mid t}\right)
\notag
\\
&=
\tilde{\alpha}_{t\mid t}
+
\mat{J}_t\left(\alpha_{t+1}-\tilde{\alpha}_{t+1\mid t}\right),
\label{eq:app_smoother_cond_mean}
\end{align}
and
\begin{align}
\Var[\alpha_t\mid \alpha_{t+1},\hat p_1,\dots,\hat p_t]
&=
\mat{P}_{t\mid t}
-
\mat{J}_t\mat{P}_{t+1\mid t}\mat{J}_t^{\top},
\label{eq:app_smoother_cond_var}
\end{align}
where
\[
\mat{J}_t=\mat{P}_{t\mid t}\mat{F}^{\top}\mat{P}_{t+1\mid t}^{-1}.
\]
\paragraph{Backward recursion for the smoothed mean.}
Taking conditional expectations in \eqref{eq:app_smoother_cond_mean} with respect to the full observed series and using the tower property gives
\begin{align}
\tilde{\alpha}_{t\mid T}
&=
\E[\alpha_t\mid \hat p_1,\dots,\hat p_T]
\notag
\\
&=
\E\!\left[
\E[\alpha_t\mid \alpha_{t+1},\hat p_1,\dots,\hat p_t]
\mid \hat p_1,\dots,\hat p_T
\right]
\notag
\\
&=
\E\!\left[
\tilde{\alpha}_{t\mid t}
+
\mat{J}_t\left(\alpha_{t+1}-\tilde{\alpha}_{t+1\mid t}\right)
\mid \hat p_1,\dots,\hat p_T
\right]
\notag
\\
&=
\tilde{\alpha}_{t\mid t}
+
\mat{J}_t
\E\!\left[
\alpha_{t+1}-\tilde{\alpha}_{t+1\mid t}
\mid \hat p_1,\dots,\hat p_T
\right]
\notag
\\
&=
\tilde{\alpha}_{t\mid t}
+
\mat{J}_t
\left\{
\E[\alpha_{t+1}\mid \hat p_1,\dots,\hat p_T]
-
\tilde{\alpha}_{t+1\mid t}
\right\}
\notag
\\
&=
\tilde{\alpha}_{t\mid t}
+
\mat{J}_t
\left(
\tilde{\alpha}_{t+1\mid T}
-
\tilde{\alpha}_{t+1\mid t}
\right).
\end{align}
The fourth equality uses that $\tilde{\alpha}_{t\mid t}$, $\tilde{\alpha}_{t+1\mid t}$, and $\mat{J}_t$ are functions of the forward Kalman filter output and are therefore fixed once the observed series is given.

\paragraph{Backward recursion for the smoothed covariance.}
Applying the law of total variance to $\alpha_t$ conditional on the full series gives
\begin{align}
\mat{P}_{t\mid T}
&=
\E\!\left[
\Var[\alpha_t\mid \alpha_{t+1},\hat p_1,\dots,\hat p_t]
\mid \hat p_1,\dots,\hat p_T
\right]
\notag
\\
&\quad
+
\Var\!\left[
\E[\alpha_t\mid \alpha_{t+1},\hat p_1,\dots,\hat p_t]
\mid \hat p_1,\dots,\hat p_T
\right].
\label{eq:app_smoother_var_ltv}
\end{align}
Substituting \eqref{eq:app_smoother_cond_mean} and \eqref{eq:app_smoother_cond_var} into \eqref{eq:app_smoother_var_ltv} yields
\begin{align}
\mat{P}_{t\mid T}
&=
\mat{P}_{t\mid t}
-
\mat{J}_t\mat{P}_{t+1\mid t}\mat{J}_t^{\top}
+
\mat{J}_t
\Var[\alpha_{t+1}\mid \hat p_1,\dots,\hat p_T]
\mat{J}_t^{\top}
\notag
\\
&=
\mat{P}_{t\mid t}
+
\mat{J}_t\left(\mat{P}_{t+1\mid T}-\mat{P}_{t+1\mid t}\right)\mat{J}_t^{\top}.
\end{align}
This establishes the smoothing covariance recursion.

\subsection{Proof of Lemma~\ref{lem:filter_smoother_variance}}
For fixed process variances, observation variances, and observation pattern, the Kalman covariance recursions do not depend on the numerical values of the observed prevalence estimates.
Consequently, $\mat{P}_{t\mid T}$ is fixed when applying the conditional variance decomposition in \eqref{eq:app_var_compare_smoother}.
Applying the conditional variance decomposition to $\alpha_t$ gives
\begin{align}
\mat{P}_{t\mid t}
&=
\Var[\alpha_t\mid \hat p_1,\dots,\hat p_t]
\notag
\\
&=
\Var\!\left\{\E[\alpha_t\mid \hat p_1,\dots,\hat p_T]\mid \hat p_1,\dots,\hat p_t\right\}
+
\E\!\left\{\Var[\alpha_t\mid \hat p_1,\dots,\hat p_T]\mid \hat p_1,\dots,\hat p_t\right\}
\notag
\\
&=
\Var\!\left\{\E[\alpha_t\mid \hat p_1,\dots,\hat p_T]\mid \hat p_1,\dots,\hat p_t\right\}
+
\mat{P}_{t\mid T}.
\label{eq:app_var_compare_smoother}
\end{align}
Taking the $(1,1)$ entry on both sides of \eqref{eq:app_var_compare_smoother} gives
\[
[\mat{P}_{t\mid t}]_{11}
=
\left[
\Var\!\left\{\E[\alpha_t\mid \hat p_1,\dots,\hat p_T]\mid \hat p_1,\dots,\hat p_t\right\}
\right]_{11}
+
[\mat{P}_{t\mid T}]_{11}.
\]
Since a covariance matrix is positive semidefinite, its $(1,1)$ entry is nonnegative. Hence
\[
\left[
\Var\!\left\{\E[\alpha_t\mid \hat p_1,\dots,\hat p_T]\mid \hat p_1,\dots,\hat p_t\right\}
\right]_{11}\ge 0,
\]
so
\[
[\mat{P}_{t\mid T}]_{11}\le [\mat{P}_{t\mid t}]_{11}.
\]
This proves Lemma~\ref{lem:filter_smoother_variance}.

\subsection{Derivation of the Innovation Likelihood}

For fixed $(Q_{\text{level}},Q_{\text{slope}})$, the Kalman filter produces $\tilde{\alpha}_{t\mid t-1}$ and $\mat{P}_{t\mid t-1}$ recursively.
From \eqref{eq:app_predictive_obs}, the conditional density of $\hat p_t$ given past observations is
\begin{equation}
f\{\hat p_t\mid \hat p_1,\dots,\hat p_{t-1}\}
=
\frac{1}{\sqrt{2\pi S_t}}
\exp\left[
-\frac{1}{2}
\frac{\{\hat p_t-\mat{H}\tilde{\alpha}_{t\mid t-1}\}^2}{S_t}
\right].
\label{eq:app_cond_density}
\end{equation}
Using the innovation notation $\nu_t=\hat p_t-\mat{H}\tilde{\alpha}_{t\mid t-1}$, this becomes
\[
f\{\hat p_t\mid \hat p_1,\dots,\hat p_{t-1}\}
=
\frac{1}{\sqrt{2\pi S_t}}
\exp\left(
-\frac{\nu_t^2}{2S_t}
\right).
\]
For any $u\le T$, let
\[
\mathcal{T}_{\mathrm{obs}}(u)=\{t:t_0<t\le u,\ \hat p_t\ \text{and}\ R_t\ \text{are observed}\}.
\]
Because the filter is initialized at $t_0$, the conditional likelihood of all subsequent observed data is obtained by the chain rule:
\begin{equation}
f\bigl(\{\hat p_t:t\in\mathcal{T}_{\mathrm{obs}}(u)\}\mid \hat p_{t_0}\bigr)
=
\prod_{t\in\mathcal{T}_{\mathrm{obs}}(u)}
f\{\hat p_t\mid \hat p_1,\dots,\hat p_{t-1}\}.
\label{eq:app_lik_factorization}
\end{equation}
Taking logarithms and substituting \eqref{eq:app_cond_density} yields
\begin{align}
\ell_u(Q_{\text{level}},Q_{\text{slope}})
&=
\sum_{t\in\mathcal{T}_{\mathrm{obs}}(u)}
\log f\{\hat p_t\mid \hat p_1,\dots,\hat p_{t-1}\}
\notag
\\
&=
\sum_{t\in\mathcal{T}_{\mathrm{obs}}(u)}
\left[
-\frac{1}{2}\log(2\pi)
-\frac{1}{2}\log S_t
-\frac{1}{2}\frac{\nu_t^2}{S_t}
\right]
\notag
\\
&=
-\frac{1}{2}
\sum_{t\in\mathcal{T}_{\mathrm{obs}}(u)}
\left\{
\log(2\pi)+\log S_t+\frac{\nu_t^2}{S_t}
\right\}.
\end{align}
The dependence on $(Q_{\text{level}},Q_{\text{slope}})$ enters through the filter recursion because both $\nu_t$ and $S_t$ depend on $\tilde{\alpha}_{t\mid t-1}$ and $\mat{P}_{t\mid t-1}$.
Setting $u=t$ and maximizing this expression yields $\widehat{\mat{Q}}_t$ for the real-time KF.
Setting $u=T$ yields the full-series process-variance estimate used for the retrospective KS.

\section{Additional Complete-Data Simulations}

\subsection{Comparisons with Simple Smoothers}
As an additional comparison, we compared the joint Kalman filter (KF) and joint Kalman smoother (KS) with simple smoothing rules applied directly to the HT series.
These comparisons were matched to the information structure of each method.
The joint KF uses $\widehat{\mat{Q}}_t$ estimated from observations through day $t$ and is therefore an online procedure, so we compare it with one-sided trailing moving averages. 
For example, a 3-day trailing moving average on day $t$ uses only $\hat p_t$, $\hat p_{t-1}$, and $\hat p_{t-2}$.
By contrast, the joint KS is a retrospective procedure that conditions on the full observed series, so we compare it with two-sided smoothers, specifically centered moving averages and LOESS. 
For instance, a 3-day centered moving average at day $t$ uses $\hat p_{t-1}$, $\hat p_t$, and $\hat p_{t+1}$, whereas the LOESS estimate at day $t$ is obtained from a local regression fit using observations from a neighborhood around $t$ on both sides.
In the missing-observation scenario, analogous moving-average or LOESS baselines would require additional ad-hoc choices about how to handle a block of missing HT values, for example through interpolation or other gap-filling rules, so we do not present those comparisons for that setting.

For the joint KF and joint KS, we use the same 95\% confidence intervals as in the main simulation study:
\[
\tilde p_{t\mid t}\pm 1.96\sqrt{[\mat{P}_{t\mid t}]_{11}}
\qquad\text{and}\qquad
\tilde p_{t\mid T}\pm 1.96\sqrt{[\mat{P}_{t\mid T}]_{11}},
\]
respectively.
For the moving-average baselines, let
\[
\hat p_t^{\mathrm{MA}}=\sum_{u\in W_t} w_{tu}\hat p_u,
\qquad
\sum_{u\in W_t} w_{tu}=1,
\]
where $W_t$ is the relevant averaging window.
To simplify the calculation, we assume independence across days, so that
\[
\Var(\hat p_t^{\mathrm{MA}})
=
\sum_{u\in W_t} w_{tu}^2 R_u,
\]
so the corresponding pointwise interval is
\[
\hat p_t^{\mathrm{MA}}\pm 1.96\sqrt{\sum_{u\in W_t} w_{tu}^2 R_u}.
\]
For the LOESS baselines, we use LOESS smoothing implemented via \texttt{stats::loess} in R and construct approximate pointwise 95\% confidence intervals using \texttt{predict.loess(}\allowbreak\texttt{..., se = TRUE)}:
\[
\hat p_t^{\mathrm{LOESS}}\pm 1.96\,\widehat{\mathrm{se}}_{\mathrm{LOESS}}(t),
\]
where $\widehat{\mathrm{se}}_{\mathrm{LOESS}}(t)$ is the model-based pointwise standard error from \texttt{predict.loess}.

\begin{figure}[h!]
    \centering
    \includegraphics[width=1\textwidth]{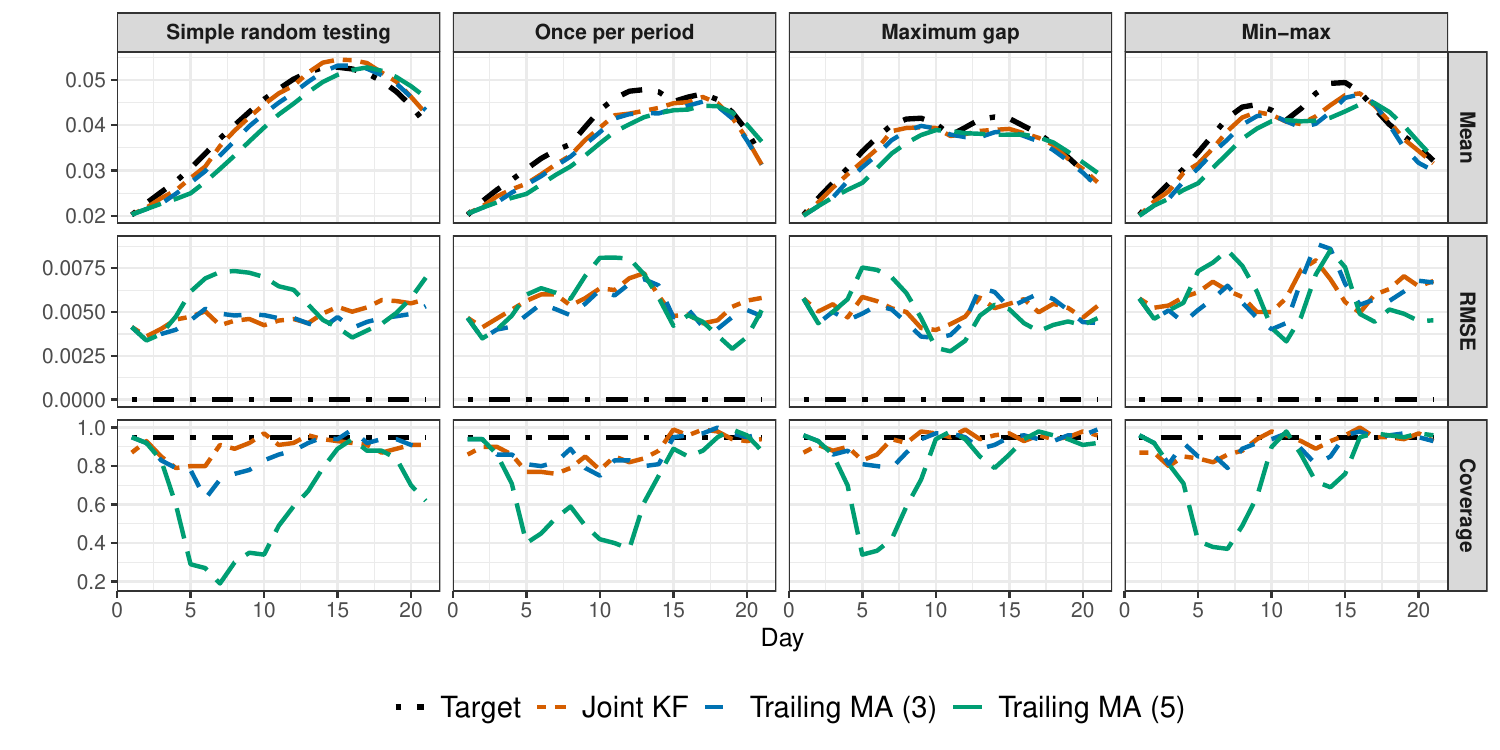}
    \caption{Complete-data simulation comparison of the real-time joint Kalman filter (KF), using $\widehat{\mat{Q}}_t$ estimated from all observations available through time $t$, with trailing moving-average baselines using 3-day and 5-day windows.
    For each testing design, the top panel shows the mean estimated prevalence over 100 simulation replicates, the middle panel shows the day-specific RMSE relative to the true prevalence, and the bottom panel shows the day-specific coverage probability of nominal 95\% confidence intervals.
    The target is shown by a black dot-dashed line, the joint KF by an orange two-dashed line, trailing MA (3) by a blue dashed line, and trailing MA (5) by a green long-dashed line.}
    \label{fig:app-kf-vs-trailing}
\end{figure}

Figure~\ref{fig:app-kf-vs-trailing} compares one-sided procedures that use only the current and past HT estimates.
Across the four testing designs, the joint KF is generally more stable over time and often shows competitive RMSE.
Its interval calibration is also better than that of the trailing moving-average baselines.
The 5-day trailing moving average tends to show the largest downward bias among these methods and is particularly unstable: its RMSE fluctuates sharply over time, and its empirical coverage can be very poor, dropping to about 0.2 under simple random testing.
The 3-day trailing moving average performs better than the 5-day version, but it still exhibits noticeable undercoverage.
A trailing moving-average analysis requires the window width to be chosen in advance, and its performance can depend on whether a 3-day or 5-day window is used.

\begin{figure}[h!]
    \centering
    \includegraphics[width=1\textwidth]{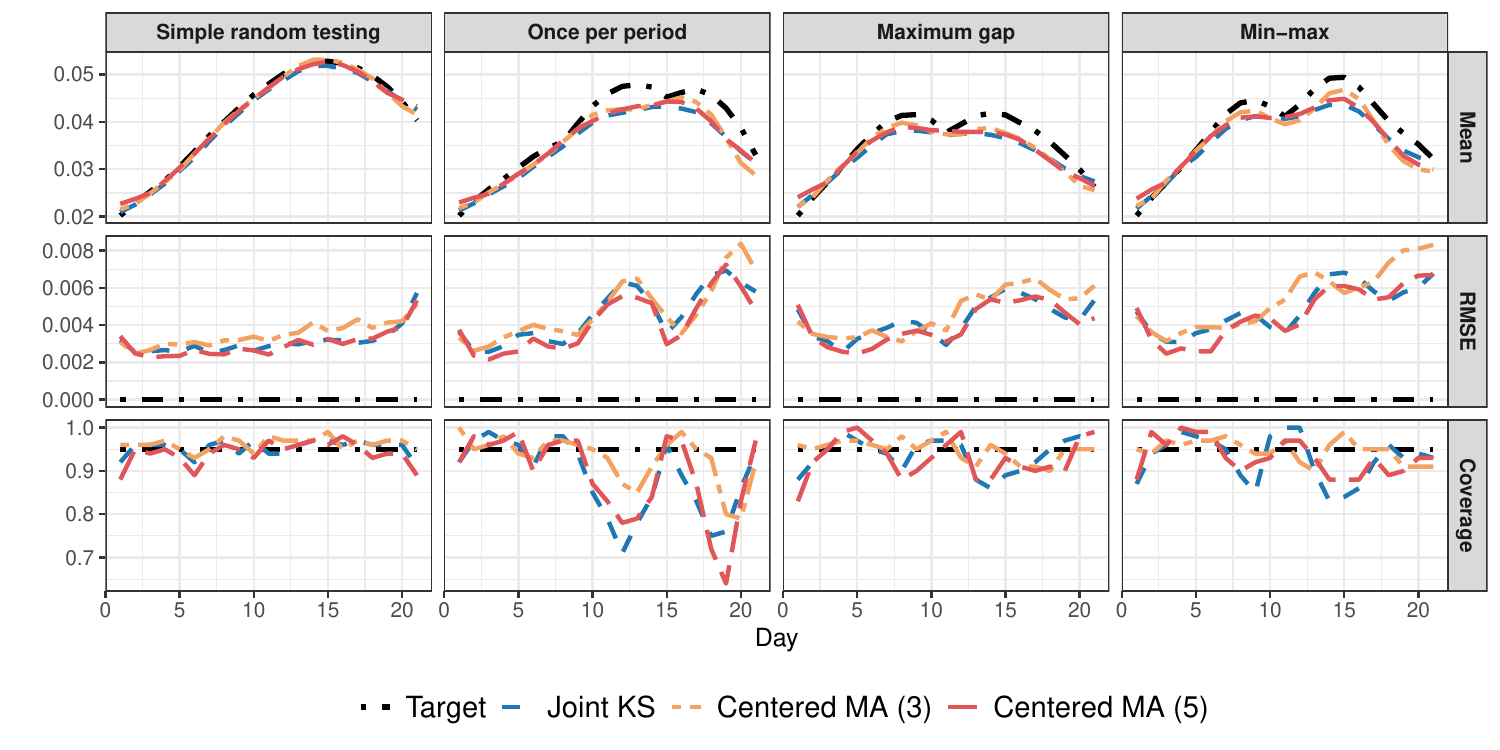}
    \caption{Complete-data simulation comparison of the retrospective joint Kalman smoother (KS), using the full-series estimate $\widehat{\mat{Q}}_T$, with centered moving-average baselines using 3-day and 5-day windows.
    The layout is the same as in Figure~\ref{fig:app-kf-vs-trailing}.
    The target is shown by a black dot-dashed line, the joint KS by a blue dashed line, centered MA (3) by an orange two-dashed line, and centered MA (5) by a red long-dashed line.}
    \label{fig:app-ks-vs-ma}
\end{figure}

Figure~\ref{fig:app-ks-vs-ma} compares the retrospective joint KS with simple symmetric averaging rules.
In this complete-data setting, the centered moving-average baselines are reasonably competitive.
The 3-day centered moving average generally has somewhat larger RMSE than the other methods, although its coverage is slightly improved relative to the joint KS in several designs.
The 5-day centered moving average is much closer to the joint KS: its RMSE is very similar overall, and its coverage is broadly comparable to, and in some cases slightly better than, that of the joint KS.
These results indicate that when the entire HT series is available, simple two-sided averaging can provide a useful retrospective reference.
However, a centered moving-average analysis still requires the window width to be chosen in advance.

\begin{figure}[h!]
    \centering
    \includegraphics[width=1\textwidth]{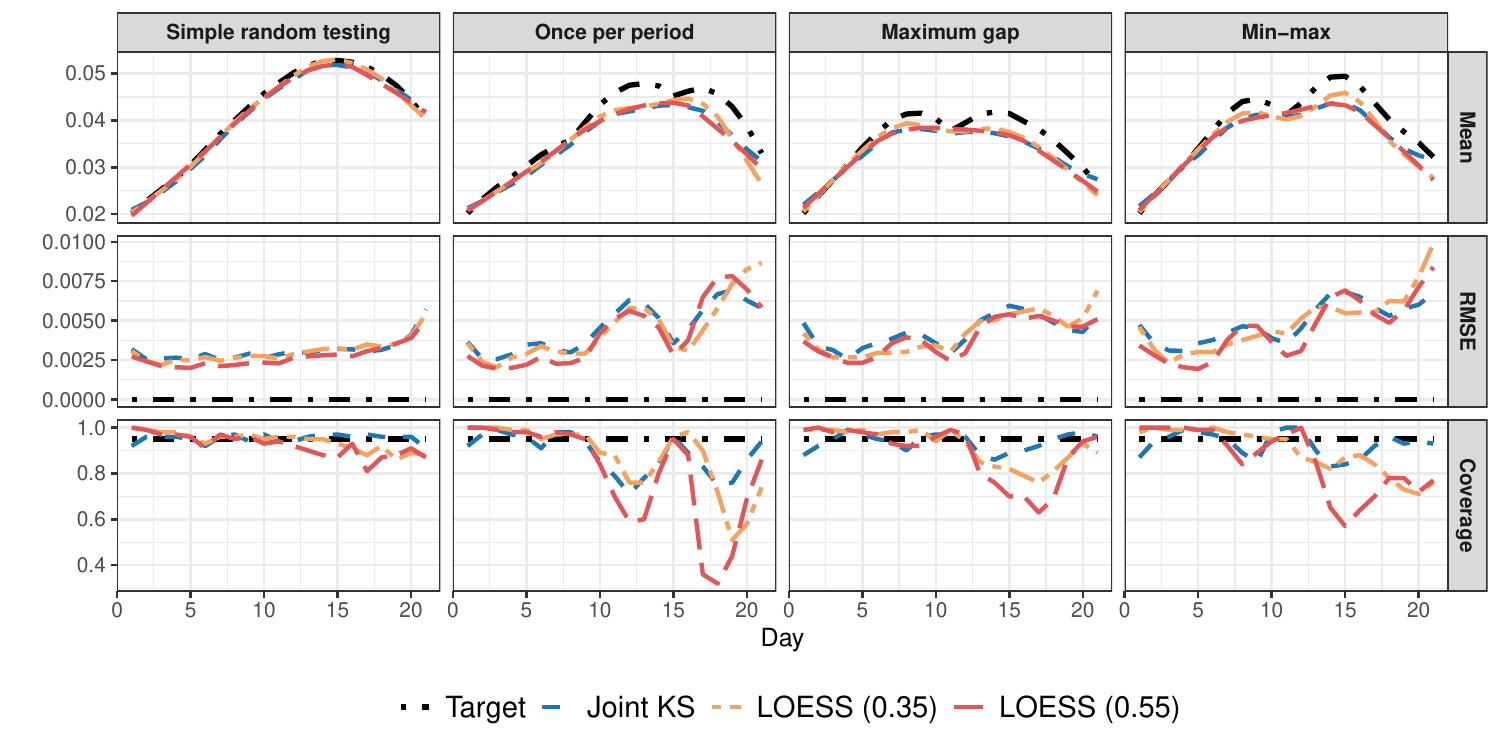}
    \caption{Complete-data simulation comparison of the retrospective joint Kalman smoother (KS), using the full-series estimate $\widehat{\mat{Q}}_T$, with unweighted LOESS baselines using spans 0.35 and 0.55.
    The layout is the same as in Figure~\ref{fig:app-kf-vs-trailing}.
    The target is shown by a black dot-dashed line, the joint KS by a blue dashed line, LOESS (0.35) by an orange two-dashed line, and LOESS (0.55) by a red long-dashed line.}
    \label{fig:app-ks-vs-loess}
\end{figure}

Figure~\ref{fig:app-ks-vs-loess} compares the joint KS with two unweighted LOESS smoothers.
The two spans represent different smoothing strengths: span 0.35 follows short-term variation more closely, whereas span 0.55 imposes heavier smoothing across nearby days.
The larger-span LOESS therefore shows slightly stronger downward bias, although the difference is not large in magnitude.
In terms of RMSE, the joint KS, LOESS (0.35), and LOESS (0.55) perform fairly similarly overall, with only modest differences among the three methods.
For interval performance, the joint KS still shows some undercoverage, but it remains closer to the nominal 95\% level than either LOESS baseline.
Because neither LOESS baseline uses the day-specific observation variances $R_t$, these comparisons should be interpreted as reference comparisons against standard nonparametric smoothers rather than as direct variance-aware competitors to the joint state-space approach.

\end{document}